\documentclass[aps,prb,twocolumn,floats]{revtex4}
\usepackage{color,ulem,verbatim,float,wrapfig}
\usepackage{amssymb,amsbsy,amsmath,mathrsfs}
\usepackage{epsfig,subfigure}
\usepackage{epstopdf}
\usepackage{graphicx}
\usepackage{times}
\usepackage{color}
\usepackage{subfigure}

\usepackage{setspace}
\usepackage{bm} % bold math.
\usepackage{caption}
\newcommand\bea{\begin{eqnarray}}
\newcommand\eea{\end{eqnarray}}
\newcommand\beq{\begin{equation}}
\newcommand\eeq{\end{equation}}
\newcommand\bib{\bibitem}

\newcommand{\noi}{\noindent}
\newcommand{\non}{\nonumber}
\newcommand{\al}{\alpha}
\newcommand{\de}{\delta}
\newcommand{\De}{\Delta}
\newcommand{\ga}{\gamma}
\newcommand{\ep}{\epsilon}

\newcommand{\lm}{\lambda}
\newcommand{\si}{\sigma}
\newcommand{\ta}{\theta}
\newcommand{\om}{\omega}
\newcommand{\dg}{\dagger}
\newcommand{\ua}{\uparrow}
\newcommand{\da}{\downarrow}
\newcommand{\pa}{\partial}
\newcommand{\la}{\langle}
\newcommand{\ra}{\rangle}

\begin{document}
%\tableofcontents

%\title{Majorana modes for a one-dimensional spin-orbit coupled Dirac system 
%with extended $s$-wave superconductivity}
\title{One-dimensional spin-orbit coupled Dirac system with extended 
$s$-wave superconductivity: Majorana modes and Josephson effects}

\author{Adithi Udupa$^1$, Abhishek Banerjee$^{2,3}$, K. Sengupta$^4$ and
Diptiman Sen$^{1,2}$}

\affiliation{$^1$Center for High Energy Physics, Indian Institute of Science,
Bengaluru 560012, India \\
$^2$Department of Physics, Indian Institute of Science, Bengaluru 560012,
India \\
$^3$Center for Quantum Devices and Microsoft Quantum Lab Copenhagen,
Niels Bohr Institute, University of Copenhagen, Universitetsparken 5,
2100 Copenhagen, Denmark \\
$^4$School of Physical Sciences, Indian Association for the Cultivation of
Science, Jadavpur, Kolkata 700032, India}

\begin{abstract}

Motivated by the spin-momentum locking of electrons at the boundaries of
certain topological insulators, we study a one-dimensional system of 
spin-orbit coupled massless Dirac electrons
with $s$-wave superconducting pairing. As a result of the spin-orbit 
coupling, our model has only two kinds of linearly dispersing modes, and we 
take these to be right-moving spin-up and left-moving spin-down. Both 
lattice and continuum models are studied. In the lattice model, we find that
a single Majorana zero energy mode appears at each end of a finite system
provided that the $s$-wave pairing has
an extended form, with the nearest-neighbor pairing being larger than
the on-site pairing. We confirm this both numerically and
analytically by calculating the winding number. We find that the
continuum model also has zero energy end modes.
Next we study a lattice version of a model with both
Schr\"odinger and Dirac-like terms and find that the
model hosts a topological transition between topologically trivial and
non-trivial phases depending on the relative strength of the
Schr\"odinger and Dirac terms. We then study a continuum
system consisting of two $s$-wave superconductors with different phases 
of the pairing, with a $\de$-function potential barrier lying at the 
junction of the two superconductors. Remarkably, we find that the system 
has a {\it single} Andreev bound state which is localized at the junction.
When the pairing phase difference crosses a multiple of $2 \pi$, an
Andreev bound state touches the top of the superconducting gap and 
disappears, and a different state appears from the bottom of the gap.
We also study the AC Josephson effect in such a junction with a voltage
bias that has both a constant $V_0$ and a term which oscillates with
a frequency $\om$. We find that, in contrast to standard
Josephson junctions, Shapiro plateaus appear when the Josephson
frequency $\om_J= 2eV_0/\hbar$ is a rational fraction of
$\om$. We discuss experiments which can realize such junctions.

\end{abstract}

\maketitle

\section{Introduction}
\label{sec1}

Topological superconductors have been studied extensively in recent years, 
largely because they have unusual states localized near the boundary of 
finite-sized systems. In particular, the Kitaev model which is a prototypical 
example of a one-dimensional topological superconductor which, in the 
topologically non-trivial phase, hosts a zero energy Majorana mode 
localized at each end of a long but finite system~\cite{kitaev1}. This is a 
lattice model in which electrons have nearest-neighbor hoppings and $p$-wave
superconducting pairing; the $p$-wave pairing implies that we can
work in a sector where all the electrons are spin polarized, and we
can therefore ignore the spin degree of freedom. The bulk spectrum of 
this system is gapped, but in the topologically non-trivial phase, each 
end hosts a localized mode whose energy lies 
in the middle of the gap with zero expectation value of the charge; these are 
the Majorana modes demonstrating fermion number fractionalization. (In contrast
to this, a model with on-site $s$-wave superconducting pairing is known not to
have such end modes). These
modes have attracted a lot of attention since an ability to braid such modes 
may eventually allow one to build logic gates and then topological quantum 
computers which are highly robust to local noise~\cite{nayak,alicea1}.

The Kitaev model and its variants have been theoretically studied in a number 
of papers~\cite{beenakker,lutchyn1,oreg,fidkowski2,potter,fulga,stanescu1,
tewari,gibertini,lim,tezuka,egger,ganga,sela,lobos,lutchyn2,cook,pedrocchi,
sticlet,jose,klinovaja,alicea2,stanescu2,asano,chung,shivamoggi,adagideli,sau1,
akhmerov,degottardi1,degottardi2,niu,sau2,lang,brouwer2,brouwer3,cai,sarkar,
chua}, and several experimental realizations have looked for the Majorana end
modes~\cite{kouwenhoven,deng,rokhinson,das,finck}. 
In addition, analytical solutions of the modes localized near the ends 
of finite-sized Kitaev chains and its generalizations have been studied 
earlier~\cite{zvyagin,kao,hegde,leumer,kawabata}.
Some common ingredients in many of the theoretical proposals and experimental 
realizations are spin-orbit coupling, an externally applied magnetic field, 
and proximity to a superconductor. 

It is known that three-dimensional topological insulators such as
Bi$_2$Se$_3$ and Bi$_2$Te$_3$ have surface states which are governed
by a massless Dirac Hamiltonian~\cite{hasan,qi}. Typically, the
Hamiltonian is given by a spin-orbit coupling term of the form
$H_{2D} = v (\si^x p_y - \si^y p_x)$, where $(p_x,p_y)$ is the
momentum of the electrons on the surface (assumed to be the $x-y$
plane here), $v$ is the velocity, and $\si^{x,y}$ denote Pauli
matrices. If we now constrict the surface to a narrow and long strip
running along the $x$-direction, the motion of the electrons in the
$y$-direction would form bands; in the lowest band, the Hamiltonian
would be given, up to a constant, by $H_{1D} = - v \si^y p_x$. Such
a model hosts a spin-dependent chirality; electrons in eigenstates
of $\si^y$ with eigenvalue $-1 ~(1)$ and $v>0$ can move only to the
right (left). (Since $\si^y$ is a good quantum number, 
if we restrict ourselves in the lowest band, we can replace
the two-component wave functions $(1,i)$ and $(1,-i)$ for $\si^y =
+1$ and $-1$ by one-component wave functions). It would then be
interesting to know what happens to this system when it is placed in
proximity to a superconductor, in particular, whether this system
can host Majorana end modes. (A similar situation would arise if we
consider a two-dimensional spin Hall insulator and look at only one
of its edges. The states at such an edge again have a spin-dependent
chirality, with the directions of the spin and the momentum being locked
to each other). We emphasize here that we are proposing to study a purely Dirac 
Hamiltonian with a spin-orbit coupled form, in contrast to the earlier models 
which generally begin with a Schr\"odinger Hamiltonian and add a spin-orbit 
term to that; it is known that the latter kind of models with combinations of 
$p$-wave and $s$-wave pairings~\cite{ray} or extended $s$-wave 
pairing~\cite{zhang1,zhang2,gaida,camjayi,aligia1,arrachea,aligia2,haim,
aksenov} 
can host Majorana end modes. These models studied earlier are known as
time-reversal-invariant topological superconductors (TRITOPS), and they have
four branches of electrons near the Fermi energy: spin-up and spin-down (or 
two channels) each of which has both right-moving and left-moving branches 
(see Ref.~\onlinecite{haim} for a review). In contrast 
to these, our model only has a right-moving spin-up and a left-moving spin-down
branch although it is still time-reversal invariant; we therefore have half 
the number of modes of a conventional TRITOPS. We will see that this leads 
to a number of unusual features, such as only one zero energy Majorana mode 
at each end of a finite system (instead of a Kramers pair of zero energy modes)
and, remarkably, only one Andreev bound state at a junction between two systems
with different superconducting pairing phases (instead of two Andreev bound
states with opposite energies). % We also wish to 
% see if Majorana modes can appear in the absence of a magnetic field.

Our study is particularly relevant in the context
of recent experimental evidence that indicates the possibility of
realizing one-dimensional Dirac-like modes at the sidewall surfaces and
crystalline edge defects of topological insulators~\cite{alpich,kandala}. 
Combined with recent encouraging
developments in fabrication of topological insulator-superconductor
heterojunctions~\cite{wang,xu,flot}, it is pertinent to understand
whether superconductivity induced into such states could produce $p$-wave 
ordering and Majorana zero modes, potentially with larger topological gaps, 
at higher sample temperatures and without an external magnetic field. 

We also study the behavior of a one-dimensional Dirac
mode in response to a superconducting phase difference induced by two
$s$-wave superconductors in a Josephson junction configuration. Josephson
junctions of $s$-wave superconductors in proximity with one-dimensional
and two-dimensional semiconductors with Rashba spin-orbit coupling
~\cite{szombati,pientka,fornieri,ren,stern}, and topological
insulators~\cite{hart,wieden}, have been extensively studied in the
context of manipulation of Majorana zero modes for topological
quantum computing, and are of immense contemporary interest.
Although Josephson junctions between a variety of quasi-one-dimensional
superconductors have been studied before~\cite{tanaka,vaccarella,kwon}, 
such junctions composed of a single one-dimensional Dirac channel have not 
been studied before. A particularly interesting Josephson effect is the 
phenomenon of Shapiro steps/plateaus. These typically appear when we 
consider a resistively and capacitively shunted Josephson junction 
in which a resistance $R$ and a capacitance $C$ are placed in parallel 
with a Josephson junction~\cite{ketterson,shukrinov,maiti,erwann,deb}.
When such a device is exposed to an external radio-frequency (rf) 
excitation, the rf drive can phase-lock with the internal dynamics of 
the Josephson junction and manifest as steps in the current versus voltage 
($I-V$) characteristics. With microwave excitation at a frequency 
$\om$, Shapiro plateaus appear as discrete plateaus in the voltage with 
quantized values $V_n=n \hbar \omega/(2e)$, where $n$ is an integer. 
In the context of topological superconductivity, the observation of missing 
plateaus at odd-integer values of $n$ has been interpreted as the 
{\it fractional} AC Josephson effect. The absence of odd-integer plateaus is 
consistent with a $4 \pi$-periodic supercurrent carried by topologically 
protected zero energy Majorana states~\cite{rokhinson,wieden,erwann}. Since 
the Josephson junctions considered in our model carry half the number of 
modes compared to the previously considered topological Josephson junctions, 
it would be interesting to understand whether the topological properties of 
our system can manifest in the AC Josephson effect.

Keeping the above considerations in mind, we have planned our paper as
follows. In Sec.~\ref{sec2}, we consider a lattice model of a system
of spin-up right-moving and spin-down left-moving Dirac electrons with 
$s$-wave superconducting pairing. We find numerically that the model has a 
topologically non-trivial phase in which there is a single zero energy 
Majorana mode at each end of a finite system, provided that the pairing is 
taken to have an extended form, and the magnitude of the 
nearest-neighbor pairing is {\it larger} than that of the on-site pairing. 
To confirm the numerical results, we present an analytical expression for 
the wave function of the end mode for a particular choice of parameters.
The different phases of the system are distinguished by a
winding number; this is zero in the topologically trivial phase and
non-zero in the topologically non-trivial phase.
We then discuss the symmetries of the model.
In Sec.~\ref{sec3}, we study a continuum version of this model.
This allows us to analytically derive the phase relation
between the two components (spin-up electron and spin-down hole) of
the wave functions for the zero energy modes at the two ends, and
this is found to be in agreement with the phase observed numerically
in the lattice model. In Sec.~\ref{sec4}, we examine a more general
model in which the Hamiltonian has both Schr\"odinger and spin-orbit
coupled Dirac-like terms. We do this since it is known that a purely 
Schr\"odinger Hamiltonian has no end modes while the spin-orbit 
coupled Dirac Hamiltonian (discussed in Sec.~\ref{sec2}) can have such 
modes; we would therefore like to see if there is a phase transition between 
the two situations.
We indeed find that a lattice version of the general model has a 
topological transition between topologically trivial and non-trivial phases 
which can be realized by tuning the relative strengths of the Schr\"odinger 
and Dirac terms. In Sec.~\ref{sec5}, we return to the spin-orbit coupled 
Dirac model with $s$-wave superconductivity and study a junction of two 
such systems with pairing phases $\phi_1$ and $\phi_2$. We find that
there is only {\it one} Andreev bound state (ABS) localized at the junction
whose energy depends on the phase difference $\De \phi = \phi_2 - \phi_1$. 
Remarkably, the ABS changes abruptly when $\De \phi$ crosses an integer 
multiple of $2 \pi$, namely, one ABS disappears after touching the top of the 
superconducting gap while another ABS appears from the bottom of the gap.
The Josephson current through the junction is however a continuous function
of $\De \phi$. [This is quite different from a standard junction 
of two $p$-wave or two $s$-wave superconductors, where there are {\it two} 
ABS with opposite energies for each value of $\De \phi$~\cite{kwon}].
We then study the AC Josephson effect in which a voltage bias
$V(t)= V_0 + V_1 \cos (\om t)$ is applied. We find multiple Shapiro plateaus 
at $\om /\om_J = m /n$, where $m, n$ are integers and $\om_J =2eV_0/\hbar$ 
is the Josephson frequency. The fact that the Josephson junction here 
exhibits Shapiro plateaus when $\om/\om_J$ is any rational fraction is in sharp
contrast to the plateaus found in generic junctions only when $\om/\om_J$ is 
an integer~\cite{ketterson,shukrinov,maiti}. Thus such plateaus distinguish 
these junctions from their standard $s$-wave counterparts. We conclude in 
Sec.~\ref{sec6} by summarizing our main results and 
discussing possible experimental realizations of our model.

Our main results can be summarized as follows. We show that a system of 
spin-orbit coupled Dirac electrons with $s$-wave superconductivity can have a 
topologically non-trivial phase where there is only one zero energy Majorana 
mode at each end. A winding number distinguishes between the topologically 
trivial and non-trivial phases. A more general model whose Hamiltonian has 
both Schr\"odinger and spin-orbit coupled Dirac terms has a phase transition
between topologically trivial and non-trivial phases depending on the 
relative strength of the Schr\"odinger and Dirac terms. A junction between
two spin-orbit coupled Dirac systems with different $s$-wave superconducting
phases with a phase difference $\De \phi$ hosts a single ABS. The application
of a voltage bias which has both a constant term $V_0$ and a term which
oscillates with frequency $\om$ shows Shapiro plateaus whenever the ratio
between $\om$ and the Josephson frequency $2eV_0 /\hbar$ is a rational number.

\section{Lattice model}
\label{sec2}

\subsection{Hamiltonian and energy spectrum}
\label{sec2a}

We consider a one-dimensional lattice system in which the electrons have
a massless spin-orbit coupled Dirac-like Hamiltonian and are in proximity
to an $s$-wave superconductor. (We will set the lattice spacing $a=1$;
hence the wave number $k$ introduced below will actually denote the
dimensionless quantity $ka$. We will also set $\hbar = 1$ unless mentioned
otherwise). The proximity-induced superconducting pairing
will be taken to have a spin-singlet form with strength
$\De_0$ for two electrons on the same site and $\De_1$ for two electrons
on nearest-neighbor sites (we will see below that the $\De_1$ term
is essential to have Majorana end modes). In terms of creation and
annihilation operators, the Hamiltonian of this lattice system has the form
\bea H_l &=& \sum_n ~[- ~\frac{i\ga}{2} ~( c_{n\ua}^{\dg}c_{n+1\ua} ~-~
c_{n+1\ua}^{\dg}c_{n\ua} ) \non \\
&& ~~~~~~~+ ~\frac{i\ga}{2} ~(c_{n\da}^{\dg}c_{n+1\da} ~-~ c_{n+1\da}^{\dg}
c_{n\da}) \non \\
&& ~~~~~~~- ~\mu ~( c_{n\ua}^{\dg}c_{n\ua} ~+~ c_{n\da}^{\dg} c_{n\da}) \non \\
&& ~~~~~~~+ ~\De_0 ~( c_{n\ua}^{\dg} c_{n\da}^{\dg} ~+~ c_{n\da} c_{n\ua})
\non \\
&& ~~~~~~~+ ~\frac{\De_1}{2} ~(c_{n\ua}^{\dg}c_{n+1\da}^{\dg} ~-~
c_{n\da}^{\dg}c_{n+1\ua}^{\dg}) \non \\
&& ~~~~~~~+ ~\frac{\De_1}{2} ~(c_{n+1\da} c_{n\ua} ~-~ c_{n+1\ua} c_{n\da})].
\label{ham1} \eea
The first two terms have the spin-orbit coupled Dirac form; in these terms, 
the signs of the hoppings (taken to be real)
is opposite for spin-up and spin-down electrons. Next,
$\mu$ denotes the chemical potential, while $\De_0$ and $\De_1$ denote on-site
and nearest-neighbor $s$-wave superconducting pairings respectively. (We have
assumed both $\De_0$ and $\De_1$ to be real. While $\De_0$ can be taken
to be real without loss of generality, we have taken $\De_1$ also to be real
for simplicity). It is convenient to replace the spin-down electron creation
(annihilation) operators with spin-up hole annihilation (creation) operators.
We will now define $c_n = c_{n\ua}$ and $d_n = 
c_{n\da}^\dg$. We then have
\bea H_l &=& \sum_n ~[-~ \frac{i\ga}{2} ~(c_n^{\dg}c_{n+1} ~-~
c_{n+1}^{\dg}c_n ) \non \\
&& ~~~~~~~+ ~\frac{i\ga}{2} ~(d_n^{\dg}d_{n+1} ~-~ d_{n+1}^{\dg}d_n) \non \\
&& ~~~~~~~- ~\mu ~( c_n^\dg c_n ~-~ d_n^\dg d_n) \non \\
&& ~~~~~~~+ ~\De_0 ~( c_n^\dg d_n ~+~ d_n^\dg c_n) \non \\
&& ~~~~~~~+ ~\frac{\De_1}{2} ~(c_n^\dg d_{n+1} ~+~ c_{n+1}^\dg d_n) \non \\
&& ~~~~~~~+ ~\frac{\De_1}{2} ~(d_{n+1}^\dg c_n+ d_n^\dg c_{n+1})]. 
\label{ham2} \eea

To find the energy spectrum of this system, we consider the equations of
motion. These are given by
\bea i \hbar \frac{dc_n}{dt} &=& [c_n, H_l] \non \\
&=& -~ \frac{i\ga}{2} ~(c_{n+1} ~-~ c_{n-1}) ~-~ \mu ~c_n \non \\
&& +~ \De_0 ~d_n ~+~ \frac{\De_1}{2} ~(d_{n+1} ~+~ d_{n-1}), \non \\
i \hbar \frac{dd_n}{dt} &=& [d_n, H] \non \\
&=& \frac{i\ga}{2} ~(d_{n+1} ~-~ d_{n-1}) ~+~ \mu ~d_n \non \\
&& +~ \De_0 ~c_n ~+~ \frac{\De_1}{2} ~(c_{n+1} ~+~ c_{n-1}). \label{eom1} \eea
Taking the second-quantized operators to be of the form
\beq c_{k,n} ~\sim~ \al e^{i (kn - Et/\hbar)} ~f_k ~~~{\rm and}~~~ d_{k,n} \sim
\beta e^{i (kn - Et/\hbar)} ~f_k, \label{cdn} \eeq
where $\al, ~\beta$ are numbers and $f_k$ is the quasiparticle 
annihilation operator for the quasiparticle with momentum $k$,
we obtain the Dirac-like eigenvalue equation
%\bea E ~\al &=& \ga \al ~\sin k ~-~ \mu \al ~+~ \De_0 ~\beta ~+~
%\De_1 \beta ~\cos k, \non \\
%E ~\beta &=& - ~\ga \beta ~\sin k ~+~ \mu \beta ~+~ \De_0 ~\al ~+~ \De_1
%\al ~\cos k. \eea
%Writing these as
%\beq \begin{pmatrix} \ga ~\sin k ~-~ \mu & \De_0 ~+~ \De_1 \cos k \\
%\De_0 ~+~ \De_1 \cos k & - (\ga ~\sin k ~-~ \mu) \end{pmatrix}
%\begin{pmatrix} \al \\ \beta \end{pmatrix} = E\begin{pmatrix}
%\al \\ \beta \end{pmatrix}. \label{ham3} \eeq
\bea && h_k ~\begin{pmatrix} \al \\ \beta \end{pmatrix} ~=~ E \begin{pmatrix}
\al \\ \beta \end{pmatrix}, \non \\
&& h_k ~=~ (\ga ~\sin k ~-~ \mu) ~\tau^z ~+~ (\De_0 ~+~ \De_1 ~\cos k) ~\tau^x,
\label{ham3} \eea 
where $\tau^{x,z}$ are Pauli matrices. This gives the energy spectrum 
\beq E ~=~ \pm \sqrt{(\ga \sin k ~-~ \mu)^{2} ~+~ (\De_0 ~+~ \De_1 
\cos k )^{2}}. \label{en1} \eeq 
We see that the gap between the positive and negative energy bands vanishes if 
$\sin k = \mu/\ga$ and $\cos k= -\De_0 /\De_1$. Hence the condition for the 
gap to close is given by
\beq \left( \frac{\mu}{\ga} \right)^{2} ~+~ \left(
\frac{\De_0}{\De_1} \right)^{2} ~=~ 1. \label{cond} \eeq
The regions $(\mu /\ga)^2 + (\De_0 /\De_1)^2 < 1$ and $> 1$ correspond 
to topologically non-trivial and trivial phases respectively. Clearly, it is 
necessary for the ratios $|\mu /\gamma |$ and $|\Delta_0 / \Delta_1|$ to be 
less than 1 in order to be in the topologically non-trivial phase.

We note here that for $\Delta_0=\Delta_1=\mu=0$, the energy 
dispersion of the quasiparticles given by Eq.\ \eqref{en1} 
% ($E_{\rm qp}=E_+-E_-$)
mimics the spectrum of the continuum Hamiltonian $H_{1D} = - v \si^y p_x$ 
with the identification $v \to \ga$ and $- \si^y \to \tau^z$. Thus one
has spin-dependent chiral electrons in the model. However, electrons of 
both chiralities are actually present in our model as must be the case with
any lattice model; namely, electrons with $k=0$ and
$k=\pi$ have opposite chiralities for a fixed $\tau^z$ and $v$. However,
as discussed at the end of Sec.~\ref{sec2b}, we can choose the parameters
$\De_0$ and $\De_1$ in such a way that the modes near $k=\pi$
do not play a significant role.

Before ending this section, we would like to show some mappings
between the momentum space Hamiltonians of our model and that of the Kitaev 
model~\cite{kitaev1}. The Hamiltonian of the Kitaev model of spin-polarized
electrons with $p$-wave superconducting pairing is given by
\bea H_K &=& \sum_n ~[- ~\frac{g_1}{2} ~(c_n^\dg c_{n+1} + c_{n+1}^\dg
c_n) ~-~ g_0 ~c_n^\dg c_n \non \\
&& ~~~~~~~+~ \frac{g_2}{2} ~(c_n^\dg c_{n+1}^\dg ~+~ c_{n+1} c_n)]. \eea
If we go to momentum space and use the basis $(c_k, c_{-k}^\dg)^T$ (where
$0 < k < \pi$), we obtain
\beq h_k ~=~ -~ (g_1 ~\cos k ~+~ g_0) ~\tau^z ~-~ g_2 ~\sin k ~\tau^y.
\label{ham4} \eeq
(The ratio $|g_0 /g_1|$ must be less than 1 to be in the topologically
non-trivial phase). It is clear that the two systems are quite different; our 
model involves both spin-up and spin-down electrons with $s$-wave 
superconducting pairing, while the Kitaev model has only a single spin (say, 
spin-up, and the spin label can therefore be ignored) and $p$-wave pairing. 
The two Hamiltonians look quite different in real space; further, our model has 
four independent parameters while the Kitaev model has three. Nevertheless, 
we find that if we set one of our parameters equal to zero, there 
are unitary transformations which relate $h_k$ in Eq.~\eqref{ham3} to the 
one in Eq.~\eqref{ham4} as follows.

\noi (i) If $\mu = 0$, $h_k$ in Eq.~\eqref{ham3} can be 
unitarily transformed to \eqref{ham4} if we change $\ga \to g_2$, $\De_0 \to 
g_0$ and $\De_1 \to g_1$.

\noi (ii) If $\De_0 = 0$, Eq.~\eqref{ham3} can be transformed to 
\eqref{ham4} if we shift the momentum $k \to k - \pi /2$ and change $\ga \to
g_1$, $\mu \to g_0$ and $\De_1 \to g_2$.

These unitary transformations imply that our model should have features 
similar to those of the Kitaev model. In particular, we will see that both 
models have only one zero energy Majorana mode at each end of a system and 
both have a winding number as a topological invariant.

\subsection{Numerical results, end modes and winding number}
\label{sec2b}

We now present numerical results for the case $\mu = 0$.
Eq.~\eqref{en1} then shows that the gap occurs when $k = 0$ or $\pi$, and
its magnitude is given by $2 | \De_1 + \De_0|$ and $2 | \De_1 - \De_0|$
respectively.

\begin{figure}[H]
\centering
\subfigure[]{\includegraphics[width= 0.9\linewidth]{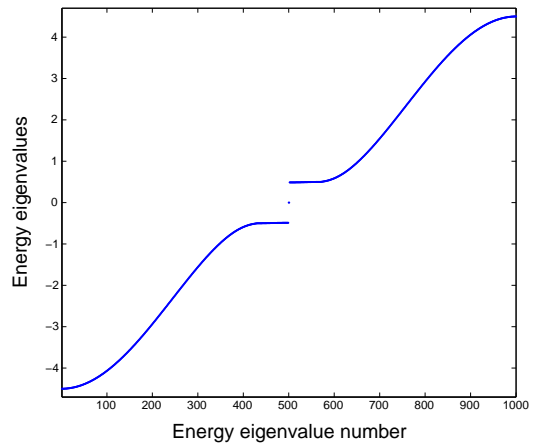}}
\subfigure[]{\includegraphics[width= 0.9\linewidth]{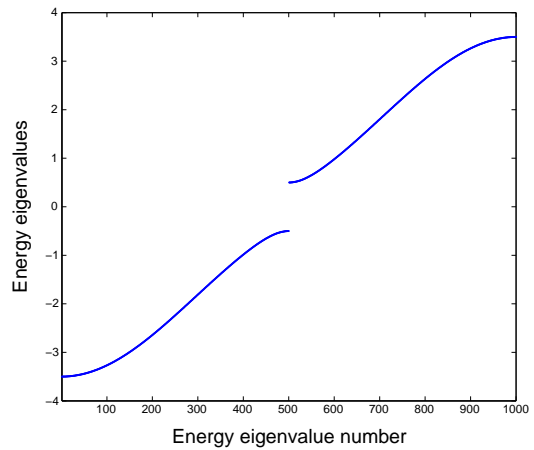}}
\caption{Energy eigenvalues for a 500-site system with $\mu=0$, $\De_0 = -2$, 
and (a) $\De_1 = 2.5$ and (b) $\De_1 = 1.5$. All energies are in units of 
$\ga$. Figure (a) shows that there are two zero energy states when $|\De_1| >
|\De_0|$, while figure (b) shows that there are no zero energy states when
$|\De_1| < |\De_0|$.} \label{fig01} \end{figure}

Numerically solving for the energy spectrum for a lattice model with
a finite number of sites and parameters $\De_0$ and $\De_1$, we find
that the energy dispersion is strikingly different in the two cases,
$|\De_0| > |\De_1|$ and $|\De_0| < |\De_1|$. We find that for
$|\De_1|< |\De_0|$, there are no states with energies lying within the 
superconducting gap. But for $|\De_1|> |\De_0|$, we find two states with zero 
energy which lie at the opposite ends of the system. This is shown in 
Fig.~\ref{fig01} for a $500$-site system with $\mu = 0$, $\De_0 = -2$, and 
$\De_1 = 2.5$ and $1.5$ (all in units of $\ga$) in figures (a) and (b) 
respectively. The $x$-axis of the figures go from 1 to 1000 since each site 
$n$ of the lattice has two variables $c_n$ and $d_n$; hence there are 1000 
states and 1000 energy levels. The energy eigenvalue number on the 
$x$-axis labels the energies in increasing order. (We have chosen a large 
enough number of sites so that if there are end modes, the hybridization 
between them is completely negligible and their energies will therefore be 
at zero energy exactly).

\begin{figure}[H]
\centering
\subfigure[]{\includegraphics[width= 0.9\linewidth]{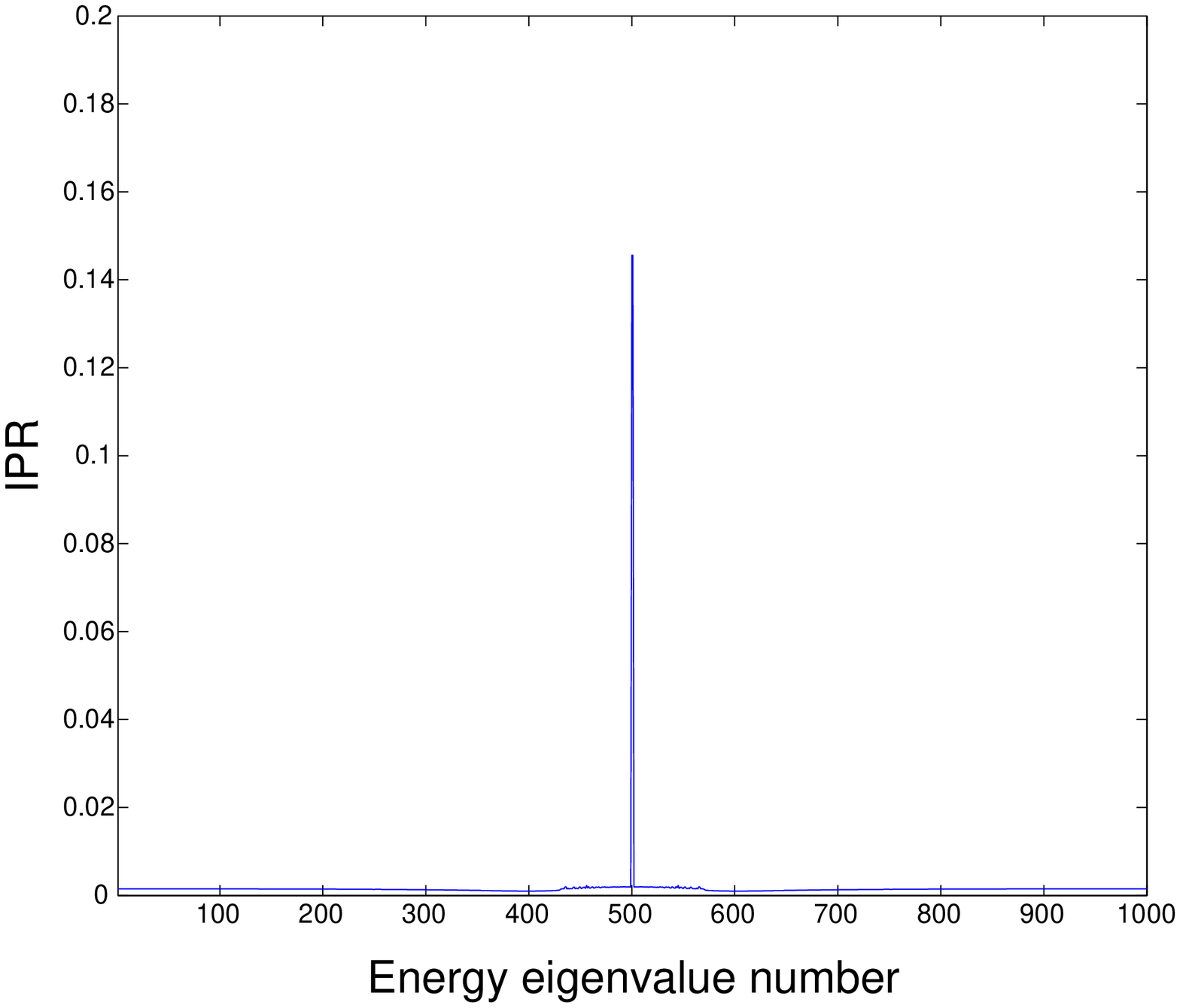}}
\subfigure[]{\includegraphics[width= 0.9\linewidth]{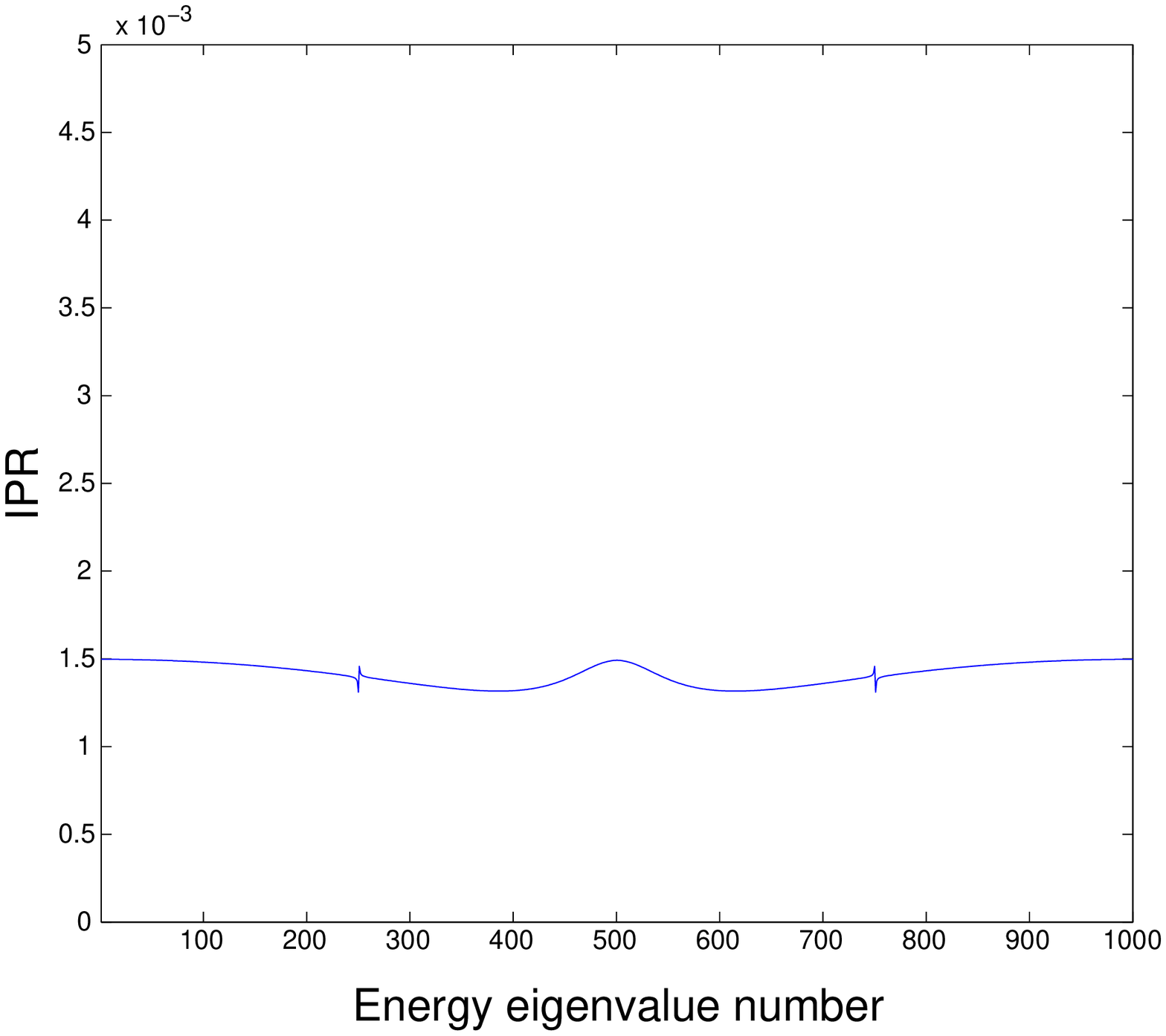}}
\caption{IPRs for a 500-site system with $\mu = 0$, $\De_0 = -2$, and 
(a) $\De_1 = 2.5$ and (b) $\De_1 = 1.5$. All energies are in units of $\ga$. 
In figure (a) where $|\De_1| > |\De_0|$, we see that two of the states have a 
much higher IPR (about $0.14$) than all the other states (which are bulk 
states); the high IPR states correspond to the end modes. In figure (b) where 
$|\De_1| < |\De_0|$, all states have approximately the same IPR (about 
$0.0015$), and they are all bulk states. (The two kinks visible in the 
figure have no physical significance).} \label{fig02} \end{figure}

To distinguish between localized and extended states, we calculate the
inverse participation ratio (IPR) calculated for all the eigenstates of the
Hamiltonian. For the $j$-eigenstate $\psi_j$, let $\psi_{j,n}$ denote
its $n$-th component, where $n$ goes from 1 to $2N$ (here $N$ is the
number of lattice sites, and the factor of 2 arises as each site has two
variables, $c_n$ and $d_n$). The IPR for $\psi_j$ is then defined as
\beq I_j ~=~ \sum_n ~|\psi_{j,n}|^4. \eeq
An extended state will generally have a value of the IPR which decreases
as the system size increases, whereas a localized state will have a finite
IPR whose value does not change with the system size. Hence a plot of the
IPR $I_j$ versus $j$ for a large system size enables us to find the localized
states easily. This is shown in Fig.~\ref{fig02} where the parameter values
have been taken to be the same as in Fig.~\ref{fig01}.

The system is said to be in a topologically non-trivial (trivial phase) if
there are end modes (no end modes) respectively. The two phases can
be distinguished from each other by a bulk
topological invariant called the winding number. Since the
Hamiltonian in Eq.~\eqref{ham3} has a form given by $H(k)= a(k)
\tau^{z} + b(k) \tau^{x}$, where $a(k) = \ga \sin k - \mu$ and $b(k)
= \De_0 + \De_1 \cos k$, we can consider a curve formed by points
given by $(a(k),b(k))$. This forms a closed curve in two dimensions
as $k$ goes from $0$ to $2\pi$. The winding number of this curve
around the origin is defined as \beq W ~=~ \frac{1}{2\pi}
\int_{0}^{2\pi} dk ~~\dfrac{a ~\pa b /\pa k ~-~ b ~\pa a /\pa k}{a^2
~+~ b^2}. \label{w} \eeq This can be evaluated numerically for
various values of $\De_0$ and $\De_1$. We find numerically that for
$|\De_1|< |\De_0|$, the winding number $W=0$ and we are in a
topologically trivial phase. For $|\De_1|> |\De_0|$, $W = \pm 1$ and we
are in a topologically non-trivial phase.
% further, $W$ is $+1$ ($-1$) if $\De_1$ is positive (negative) respectively.

\begin{figure}[H]
\centering
\subfigure[]{\includegraphics[width= 0.9\linewidth]{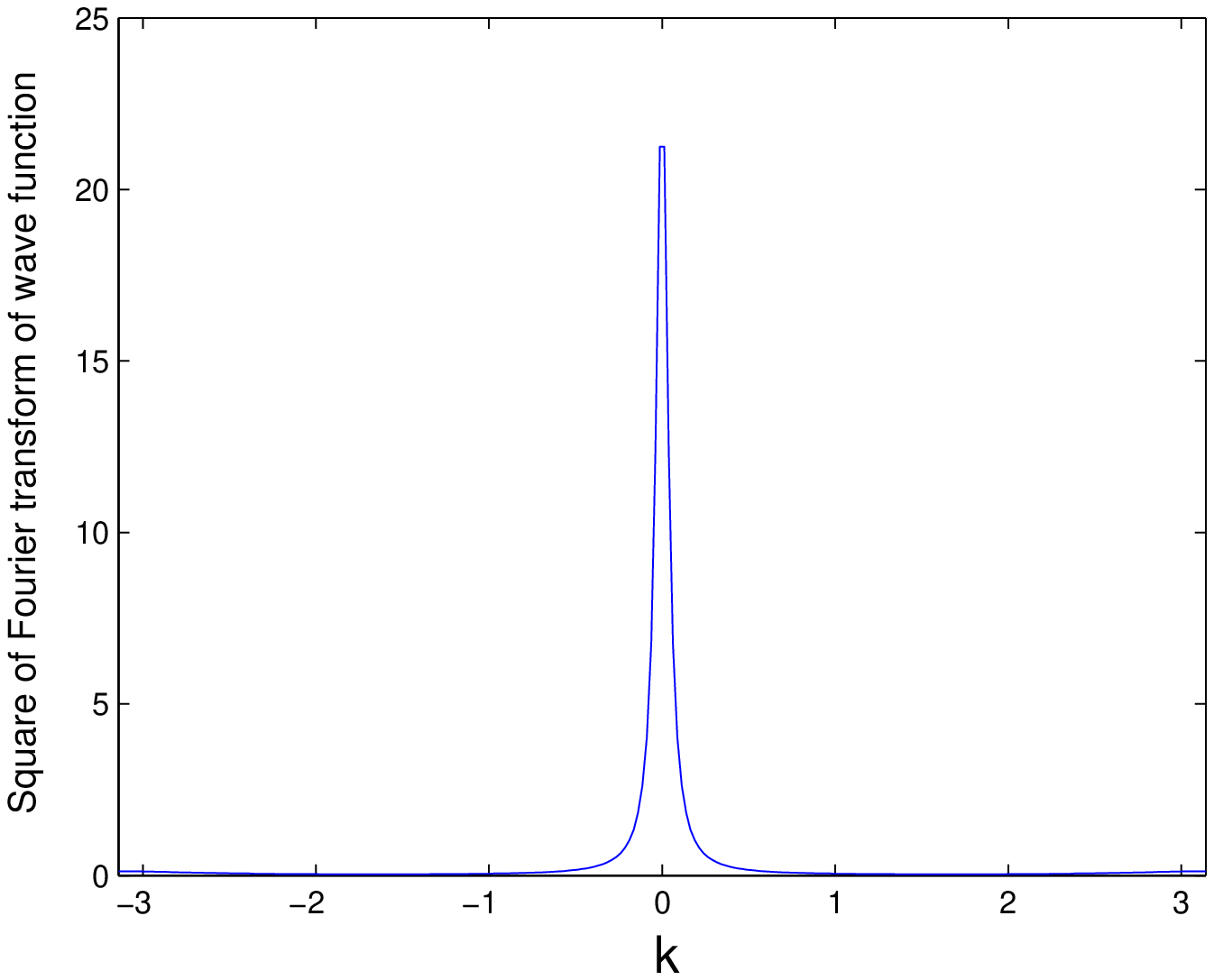}}
\subfigure[]{\includegraphics[width= 0.9\linewidth]{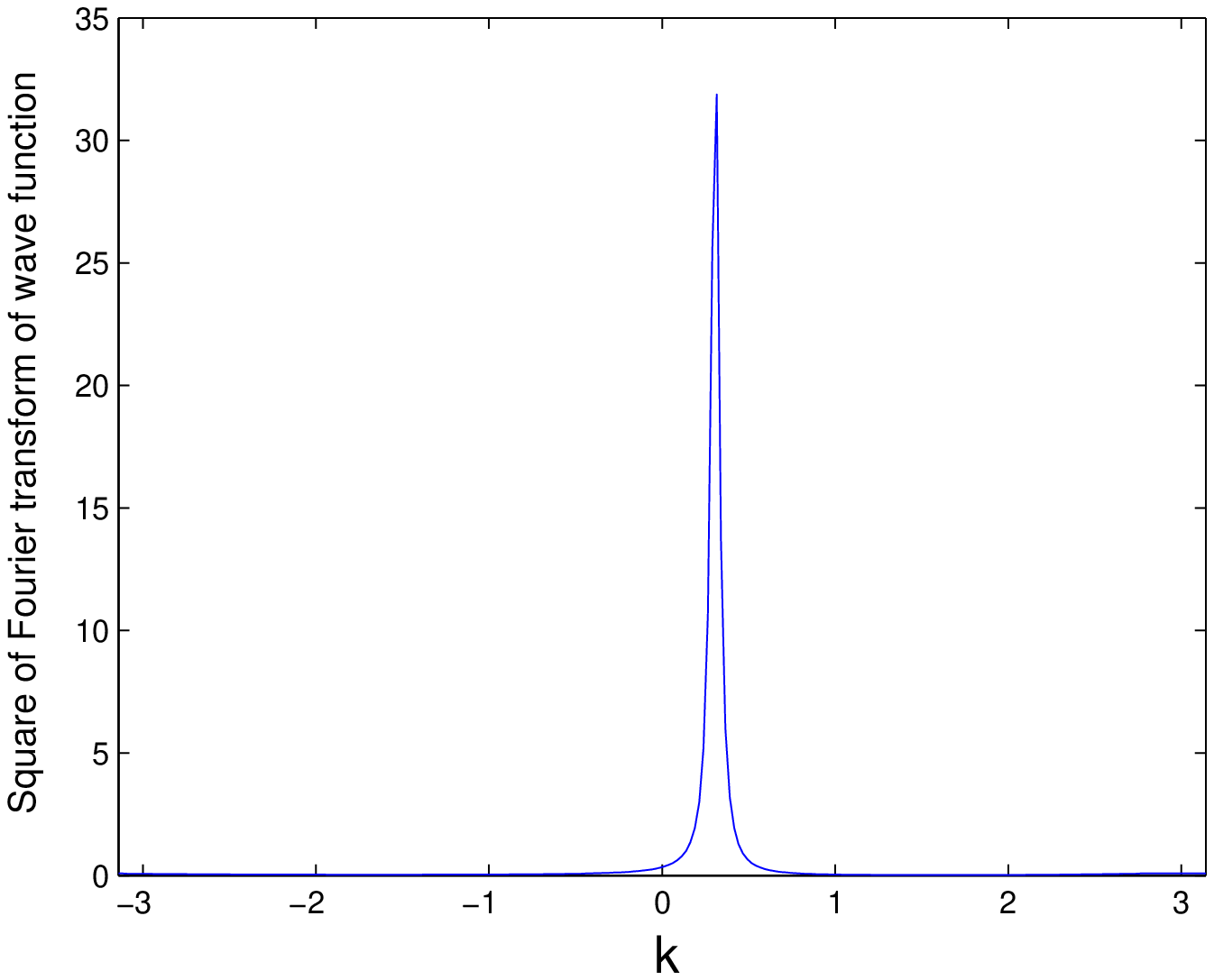}}
\caption{Absolute value squared of the Fourier transform, 
$|{\tilde c}_k|^2 + |{\tilde d}_k|^2$,
of the wave function of the mode at the left end of a 500-site system with 
$\De_0 = -0.26$ and $\De_1 = 0.3$. All energies are in units of $\ga$.
In (a), $\mu = 0$ and the Fourier transform has a peak at $k=0$. In (b),
$\mu = 0.3$ and the Fourier transform has a peak at $k=0.3$. In both cases,
the peak width at half maximum is about $0.08$.} \label{fig03} \end{figure}

It is instructive to look at the Fourier transforms of the wave functions of
the end modes. Given the wave function $(c_n,d_n)$ of an end mode, we
calculate the Fourier transforms $({\tilde c}_k, {\tilde d}_k)$, and plot
$|{\tilde c}_k|^2 + |{\tilde d}_k|^2$ versus $k$. This is shown in
Fig.~\ref{fig03} for a 500-site system with $\De_0 = -0.26$ and $\De_1 = 0.3$ 
in units of $\ga$; we have taken $\mu = 0$ in Fig.~\ref{fig03} (a) and
$\mu = 0.3$ in Fig.~\ref{fig03} (b). The locations and widths of the peaks in
the two figures can be understood as follows. Since the end mode has zero
energy, Eq.~\eqref{en1} implies that the momentum $k$ should satisfy
\beq (\ga \sin k ~-~ \mu)^2 ~+~ (\De_0 ~+~ \De_1 \cos k )^2 ~=~ 0.
\label{f} \eeq
For $\mu, ~\De_0, ~\De_1 \ll \ga$, the solution of Eq.~\eqref{f} is given by
\beq k ~\simeq~ \frac{\mu ~\pm~ i ~|\De_0 ~+~ \De_1|}{\ga}. \eeq
For the mode at the left end, the wave function $c_n, ~d_n \sim e^{ikn}$
should have the imaginary part of $k$ positive so that the wave function goes
to zero as $n \to + \infty$. Hence we must take $k = (\mu + i |\De_0 + \De_1|)/
\ga$, implying that the wave function goes as $e^{in (\mu + i |\De_0 +
\De_1|)/\ga}$. The Fourier transform of this has a peak at $k=\mu /\ga$ and
a width equal to $2 |\De_0 + \De_1|/\ga$. This agrees with the locations
and widths of the peaks that we see in Fig.~\ref{fig03}.

% figures generated by test31.m

We would like to emphasize here that our model has only one zero energy
mode at each end of a long system, in contrast to conventional TRITOPS which 
have a Kramers pair of zero energy modes at each 
end~\cite{zhang1,zhang2,gaida,camjayi,aligia1,arrachea,aligia2,haim,aksenov}. 
While we have shown this numerically above, we can also show this analytically 
for the special case corresponding to $\mu = \De_0 = 0$. We will look for zero 
energy modes localized near the left end of a semi-infinite system where the 
sites go as $n=0,1, 2,\cdots$. For zero energy, the left hand sides of 
Eqs.~\eqref{eom1} vanish; we then obtain the recursion relation
\beq \left( \begin{array}{c}
c_{n+2} \\
d_{n+2} \end{array} \right) ~=~ \frac{1}{\ga^2 ~-~ \De_1^2} ~
\left( \begin{array}{cc}
\ga^2 ~+~ \De_1^2 & -i 2 \ga \De_1 \\
i 2 \ga \De_1 & \ga^2 ~+~ \De_1^2 \end{array} \right) ~\left( \begin{array}{c}
c_n \\
d_n \end{array} \right) \label{eqn1} \eeq
for $n \ge 1$, and 
\bea - i \ga c_1 ~+~ \De_1 d_1 &=& 0, \non \\
i \ga d_1 ~+~ \De_1 c_1 &=& 0. \label{eqn2} \eea
If $\De_1 \ne \pm \ga$, Eq.~\eqref{eqn2} gives $c_1 = d_1 = 0$;
Eq.~\eqref{eqn1} then implies that $c_n = d_n = 0$ for all odd values of
$n$. Next, the eigenvalues of the matrix appearing on the right hand side
of Eq.~\eqref{eqn1} are given by $(\ga + \De_1)/(\ga - \De_1)$ and
$(\ga - \De_1)/(\ga + \De_1)$. If $\ga$ and $\De_1$ have the same sign
the first eigenvalue is larger than 1 while the second eigenvalue is smaller
than 1. To get a normalizable state, we must choose $(c_n,d_n)^T$ to be the
eigenstate corresponding to the second eigenvalue. This implies the unique
solution
\beq \left( \begin{array}{c}
c_n \\
d_n \end{array} \right) ~=~ \left( \frac{\ga ~-~ \De_1}{\ga ~+~ \De_1} 
\right)^{n/2}~ \left( \begin{array}{c}
1 \\
-i \end{array} \right), \eeq
up to an overall normalization, where $n=0,2,4,\cdots$. We have confirmed
that this matches the numerical results for the zero energy mode at the 
left end of the system. (We note that our analytical solution for the
end mode of a semi-infinite system has some similarities with the solutions 
discussed earlier for the end modes of a finite-sized Kitaev chain and some 
of its generalizations~\cite{zvyagin,kao,hegde,leumer,kawabata}).

Before ending this section, we would like to comment on the fermion doubling
problem which generally plagues lattice models with a massless
Dirac Hamiltonian~\cite{nielsen} and
which does not appear in continuum models such as the one discussed in the next
section. For instance, if we set $\mu = \De_0 = \De_1 = 0$ in Eq.~\eqref{ham3},
the energy given by $\pm \ga \sin k$ vanishes at both $k=0$ (which has a smooth
continuum limit) and $k=\pi$ (which does not have a smooth continuum limit).
We may therefore worry that the end modes that we have found numerically may
be artefacts of the lattice model and more specifically of fermion doubling.
However, we find numerically that this is not so. If we choose $\De_0$ and
$\De_1$ to have opposite signs and close to each other in magnitude, and
$\mu = 0$, we see from Eq.~\eqref{en1} that the superconducting gap vanishes
at $k=0$ and not $k=\pi$. We then find that the absolute value squared of 
the Fourier transform, $|{\tilde c}_k|^2 + |{\tilde d}_k|^2$, of the 
wave function of the end modes is much larger around $k=0$ than around 
$k= \pi$ (see Fig.~\ref{fig03}). Thus the doubled modes appearing near $k=\pi$ 
do not contribute significantly to the end modes. Further, we will see in 
Sec.~\ref{sec3} that the continuum model also has end modes, confirming that 
the modes near $k=0$ of the lattice model have a smooth continuum limit.

\subsection{Symmetries of the model}
\label{sec2c}

We would now like to discuss some of the symmetries of our model. 
It is convenient to separately discuss symmetries of the Schr\"odinger 
equations in Eqs.~\eqref{eom1} which act on wave functions and symmetries
of the Hamiltonian in Eq.~\eqref{ham1} which act on second-quantized 
operators.

We find that Eqs.~\eqref{eom1} have the following two symmetries.

\noi (i) Combination of $t \to - t$ and particle-hole symmetry:
Eqs.~\eqref{eom1} remain the same if we change $t \to - t$, $c_n \to d_n$ and
$d_n \to - c_n$. (Note that we do {\it not} do complex conjugation). Hence, if
there is a solution with wave function $(c_n,d_n)$ and energy $E$, there will
also be a solution with wave function $(d_n,-c_n)$ and the opposite energy
$-E$ (since $e^{-iEt/\hbar} \to e^{iEt/\hbar}$ under $t \to - t$).
The combination of these two symmetries, called chiral 
symmetry~\cite{schnyder}, explains why we have a winding number as a 
topological invariant.

\noi (ii) Combination of complex conjugation, $t \to - t$ and parity:
Eqs.~\eqref{eom1} remain the same if we complex conjugate them, change $t \to
-t$, and invert $n \to -n$. This implies that if there is a solution with wave
function $(c_n,d_n)$ and energy $E$, there will also be a solution with wave
function $(c^*_{-n},d^*_{-n})$ and the same energy $E$ (since $e^{-iEt/\hbar}$
remains the same under complex conjugation and $t \to - t$).

These symmetries imply that if we take a finite-sized system and there is
only one mode localized at, say, the left end, then its energy must be equal
to zero (due to symmetry (i)), and there must also be a zero energy mode
localized at the right end (due to symmetry (ii)). These agree with the
numerical results presented in Sec.~\ref{sec2b}. 
% The existence of these symmetries along with the existence of a winding 
% number which is a $Z$-valued topological invariant enable us to identify 
% the symmetry class of this system as 
% BDI~\cite{schnyder,teo,fidkowski1,degottardi2}.

Symmetry (i) also implies that if there is a zero energy mode at one
end of a system, the expectation value of the charge in that mode,
given by \beq Q ~=~ -e ~\sum_n ~(|c_n|^2 - |d_n|^2) \label{charge}
\eeq (where $-e$ is the electron charge), must be invariant under
$c_n \to d_n$ and $d_n \to - c_n$, and must therefore be equal to zero. 

We now discuss the symmetries of the Hamiltonian in Eq.~\eqref{ham1}.
We find that there are two antiunitary symmetries.

\noi (i) Time-reversal, i.e., spin-reversal and complex 
conjugation: Eq.~\eqref{ham1} remains the same if we transform $c_{n \ua} \to 
c_{n \da}$, $c_{n \da} \to - c_{n \ua}$, and do complex conjugation. We note 
that the square of this transformation is equal to $-1$. The existence of 
this symmetry implies that the symmetry class of this system is 
DIII~\cite{schnyder,teo,fidkowski1}. 
%However, we have seen at the end of Sec.~\ref{sec2a} that the momentum-space 
%Hamiltonian of our model can be unitarily transformed to that of the Kitaev 
%model, and it is known that the Kitaev model lies in the class BDI and has 
%a $Z$-valued invariant~\cite{degottardi2}.
%Note that although a model in the symmetry class DIII should have a 
%$Z_2$ invariant (which we will not consider here), our model 
%has a $Z$-valued invariant as we have discussed above.

\noi (ii) Combination of complex conjugation and parity: Eq.~\eqref{ham1} 
remains the same if we transform $c_{n \ua} \to c_{-n \ua}$, $c_{n \da} \to 
c_{-n \da}$, and do complex conjugation. The square of this transformation 
is $+1$.

The symmetries discussed above can be broken in a variety of ways. A
simple example is given by the case where the on-site
superconducting pairing $\De_0$ is complex, so that the
corresponding terms in Eq.~\eqref{ham1} are given by $\De_0
c_{n\ua}^{\dg} c_{n\da}^{\dg} + \De^*_0 c_{n\da} c_{n\ua})$. We then
find that both the symmetries given above are broken, although the
combination of the two is still a symmetry (i.e., complex conjugate
Eqs.~\eqref{eom1} and change $c_n \to d^*_{-n}$ and $d_n \to
-c^*_{-n}$), implying that if there is a mode at the left end with
energy $E$, there will be a mode at the right end with energy $-E$.
Numerically, we indeed find that if $\De_0$ is complex, the modes at
the right and left ends generally have energies $E$ and $-E$
respectively, where $E \ne 0$. Further, Eq.~\eqref{ham3} has an
additional term given by ${\rm Im} (\De_0) \tau^y$. Hence the
Hamiltonian now has a combination of three Pauli matrices, i.e.,
Hamiltonian $H(k) = a(k) \tau^z + b(k) \tau^x + e(k) \tau^y$. As a
function of $k$, $(a(k),b(k),e(k))$ defines a closed curve in three
dimensions, instead of two dimensions. Hence it is no longer
possible to define a winding number.

We can analytically find the energies of the end modes when $\De_0$ is
complex as follows. We first take $\De_0$ to be real. We then know that
Eqs.~\eqref{eom1}, which we can re-write as
\bea && -~ \frac{i\ga}{2} ~(c_{n+1} ~-~ c_{n-1}) ~-~ \mu ~c_n \non \\
&& +~ \De_0 ~d_n ~+~ \frac{\De_1}{2} ~(d_{n+1} ~+~ d_{n-1}) ~=~ E c_n, \non \\
&& \frac{i\ga}{2} ~(d_{n+1} ~-~ d_{n-1}) ~+~ \mu ~d_n \non \\
&& +~ \De_0 ~c_n ~+~ \frac{\De_1}{2} ~(c_{n+1} ~+~ c_{n-1}) ~=~ E d_n,
\label{eom2} \eea
has solutions at the ends with $E=0$. Further, we will see in Sec.~\ref{sec3}
that if $\ga > 0$, the mode at the left end has $d_n = - i c_n$ while the
mode at the right end has $d_n = i c_n$. This implies that for $E=0$, the
two equations in \eqref{eom2} reduce to the equations
\bea && -~ \frac{i\ga}{2} ~(c_{n+1} ~-~ c_{n-1}) ~-~ \mu ~c_n \non \\
&& \mp ~i ~\De_0 ~c_n ~\mp~ i ~\frac{\De_1}{2} ~(c_{n+1} ~+~ c_{n-1}) ~=~ 0,
\non \\
&& \frac{i\ga}{2} ~(d_{n+1} ~-~ d_{n-1}) ~+~ \mu ~d_n \non \\
&& \pm ~i ~\De_0 ~d_n ~\pm~ i ~\frac{\De_1}{2} ~(c_{n+1} ~+~ c_{n-1}) ~=~ 0,
\label{eom2b} \eea
where the upper (lower) signs in both the equations hold for the left (right)
end modes respectively. Now, suppose that $\De_0$ is complex; let us denote it
by ${\tilde \De}_0$ to distinguish it from the real $\De_0$ in
Eq.~\eqref{eom2b}. Since the modes at the left (right) ends
satisfy $d_n = \mp i c_n$ respectively, we obtain the equations
\bea && -~ \frac{i\ga}{2} ~(c_{n+1} ~-~ c_{n-1}) ~-~ \mu ~c_n \non \\
&& \mp ~i ~{\tilde \De}_0 ~c_n ~\mp~ i ~\frac{\De_1}{2} ~(c_{n+1} ~+~ c_{n-1})
~=~ E c_n \non \\
&& \frac{i\ga}{2} ~(d_{n+1} ~-~ d_{n-1}) ~+~ \mu ~d_n \non \\
&& \pm ~i ~{\tilde \De}^*_0 ~d_n ~\pm~ i ~\frac{\De_1}{2} ~(c_{n+1} ~+~
c_{n-1}) ~=~ E d_n. \label{eom2c} \eea
We now observe that Eqs.~\eqref{eom2c} can be mapped to Eqs.~\eqref{eom2b}
if we replace ${\rm Re} ({\tilde \De}_0) \to \De_0$ and $E \to \pm {\rm Im}
({\tilde \De}_0)$, where the $\pm$ hold for the left (right) end modes
respectively. We thus conclude that when the on-site pairing $\De_0$ becomes
complex, the wave functions $(c_n,d_n)$ of the end modes do not change
(if we do not change the value of ${\rm Re} (\De_0$)), but their energies
change from zero to $\pm {\rm Im} (\De_0)$ at the left (right) ends
respectively. Interestingly, the fact that the wave functions of the end modes
do not change when $\De_0$ becomes complex implies that the expectation
values of the charge (defined in Eq.~\eqref{charge}) remain equal to
zero even though their energies become non-zero. 
%Hence the wave functions retain their Majorana character.

\begin{figure}[H]
\centering
\subfigure[]{\includegraphics[width= 0.9\linewidth]{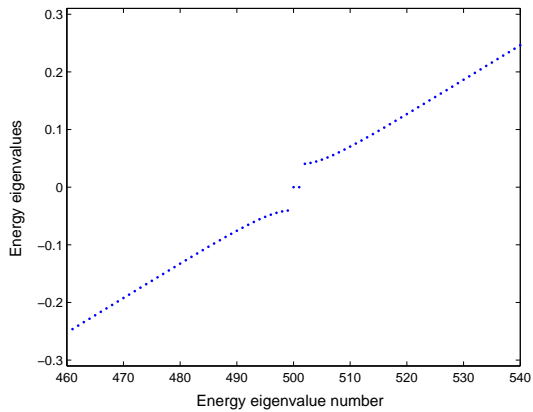}}
\subfigure[]{\includegraphics[width= 0.9\linewidth]{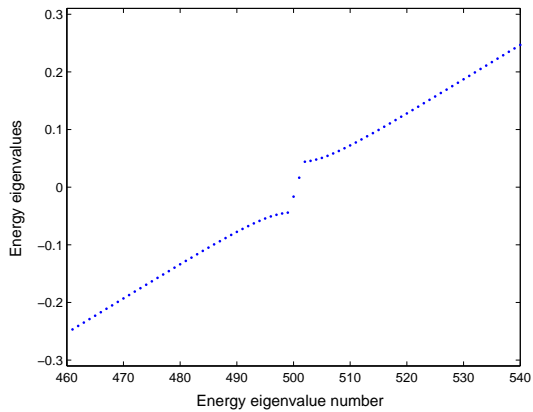}}
\caption{Enlarged view of energy eigenvalues close to zero for a 500-site 
system with $\mu = 0$, and $\De_1 = 0.3$. In (a), $\De_0 = -0.26$ is real and 
each end has a zero energy mode. The superconducting gap is $0.081$. In (b), 
$\De_0 = -0.26 e^{i \pi/50}$ is complex, and the left (right) end has a mode 
with energy $-0.0163$ ($+0.0163$) respectively. The superconducting gap is 
$0.088$. All energies are in units of $\ga$.} \label{fig04} \end{figure}

Figures~\ref{fig04} (a) and (b) show the effect of symmetry breaking on the 
end mode energies of a $500$-site system with $\mu = 0$ and $\De_1 = 0.3$. 
In Fig.~\ref{fig04} (a), $\De_0 = -0.26$ is real and each end has a zero 
energy mode. In Fig.~\ref{fig04} (b), $\De_0 = -0.26 e^{i \pi/50}$ is complex,
and the end modes have energies $-0.0163$ (left end) and $0.0163$ (right end).
We note that these values agree with $\pm {\rm Im} (\De_0)$ respectively. 
(All energies are in units of $\ga$).

% figures generated by test31.m

\section{Continuum model}
\label{sec3}

We now consider a continuum model for a system with spin-orbit coupled
Dirac Hamiltonian and an $s$-wave superconducting pairing which is a complex
number. The continuum Hamiltonian is given by
\bea H_c &=& \int dx ~[ - ~i \ga ~(c^\dg \pa_{x} c ~-~ d^\dg \pa_{x}
d) \non \\
&& ~~~~~~~~~~+ \De ~e^{i \phi} ~c^\dg d ~+~ \De ~e^{-i\phi} ~d^\dg c
], \label{ham5} \eea where $\ga$ denotes the velocity. (We have
assumed $\mu = 0$ for simplicity). Note that the unlike the lattice
model which has two different pairing parameters $\De_0$ and
$\De_1$, a continuum model can only have one parameter $\De$. We saw
in Sec.~\ref{sec2} that if $\De_0$ and $\De_1$ have opposite signs
and are close to each other in magnitude, the long-distance
properties of the lattice model are dominated by modes with momenta
close to $k=0$. The form of the Hamiltonian in Eq.~\eqref{ham3} then
implies that the pairing $\De$ in the continuum model is related to
the pairings $\De_0, ~\De_1$ in the lattice model as \beq \De ~=~
\De_0 ~+~ \De_1. \eeq

Assuming the form in Eq.~\eqref{cdn}, Eq.~\eqref{ham5} leads to the equation
\beq \begin{pmatrix}
-i\ga \pa_{x} & \De e^{i\phi} \\
\De e^{-i\phi} & i\ga \pa_{x}
\end{pmatrix} \begin{pmatrix} \al \\ \beta
\end{pmatrix} ~=~ E\begin{pmatrix} \al \\ \beta
\end{pmatrix}. \label{eom3} \eeq
This gives the energy spectrum
\beq E ~=~ \pm \sqrt{\De^2 + \ga^2 k^2}. \label{en2} \eeq
This has a gap from $-\De$ to $+\De$. In the rest of this section, we will
set $\phi = 0$ and $\De > 0$. This can be done without loss of generality
since we can absorb the phase $e^{i\phi}$ in $d$ in Eq.~\eqref{ham5}.

\begin{figure}[H]
\centering
\includegraphics[scale=0.52]{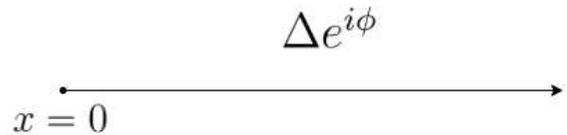}
\caption{Schematic picture of a semi-infinite system terminated on the left at
$x=0$. The $s$-wave pairing $\De e^{i \phi}$ is indicated.}
\label{fig05} \end{figure}

This system has the same symmetries as discussed in Sec.~\ref{sec2c}. Namely,
the equations of motion remain invariant under under (i) $t \to - t$, 
$c(x) \to d(x)$, and $d(x) \to - c(x)$),
and (ii) $t \to -t$, $c(x) \to c^*(-x)$, and $d(x) \to d^* (-x)$).

To study a localized mode which can appear at one end of the system, we now
consider a semi-infinite system which is terminated at the left end, at $x=0$.
The system goes from $x=0$ to $\infty$ as indicated in Fig.~\ref{fig05}. To
obtain a localized state whose energy lies within the superconducting gap,
$-\De < E< \De$, we require a wave function which decays as $x$ increases.
Hence the wave number $k$ appearing in Eq.~\eqref{en2} must have the form
$k = (i/|\ga|) \sqrt{\De^2 - E^2}$.

Next, we impose the condition that the probability current $J$ must be zero at
$x=0$. We can derive an expression for $J$ by defining the probability density
$\rho = c^\dg c + d^\dg d$ and demanding that the equations of motion must
lead to the equation of continuity, $\pa_t \rho + \pa_x J = 0$. This gives
$J = \ga (c^\dg c - d^\dg d)$. We must therefore have $c^\dg c - d^\dg d= 0$
at $x=0$. The general solution to this is $d= e^{i\ta}c$, where $\ta$ can be
an arbitrary real parameter. However, the symmetry (i) mentioned above implies
that we must have $e^{i\ta} = \pm i$, i.e., $\ta = \pm \pi/2$. Substituting
this in the equations of motion, we obtain
\beq \dfrac{E ~-~ i ~{\rm sgn} (\ga) \sqrt{\De^2 -E^2}}{\De} ~=~ e^{i\ta}, \eeq
where $\ta$ is $-\pi/2$ if $\ga > 0$ and $\pi/2$ if $\ga < 0$, and ${\rm sgn}
(\ga)$ denotes the sign of $\ga$. In either case, we have $E = 0$.

Similarly, for a system terminated at the right end, with $x$ decreasing as we
go away from the end and into the system, we find that we must choose
$k =- (i/|\ga|) \sqrt{\De^2 - E^2}$. We now find that the allowed values of
$\ta$ are $\pi/2$ if $\ga > 0$ and $-\pi/2$ if $\ga < 0$.

These conditions on $\ta$ give the relation between the two components of the
wave function as $\beta = \mp i \al$ at the left (right) ends respectively,
if $\ga > 0$. We find numerically that the end modes of the lattice model
indeed have $E=0$ and their wave functions satisfy the relations given above.
We note here that the phase relation between the two components holds
for all values of $x$, not just at the two ends. Namely, the mode localized at
the left (right) end has $\beta (x) = \mp i \al (x)$ for all $x$.

\section{General model with both Dirac and Schr\"odinger terms}
\label{sec4}

In this section we will consider a more general model in which the
Hamiltonian is a combination of a spin-orbit coupled Dirac
Hamiltonian, a Schr\"odinger Hamiltonian, and an $s$-wave
superconducting pairing. The motivation for this study is as
follows. We know that in the presence of $s$-wave superconducting
pairing, a purely Schr\"odinger Hamiltonian without a spin-orbit
coupling term has no zero energy end modes, while a purely Dirac
Hamiltonian with a spin-orbit coupled form does have such modes. We
therefore want to know how a transition between the two phases
occurs when going from one limit to the other.

We will take the total continuum Hamiltonian to be 
\bea H_c &=& \int dx ~[ - ~i \ga ~(c^\dg \pa_{x} c ~-~ d^\dg \pa_{x}
d) \non \\
&& ~~~~~~~~~~- ~\frac{\ep \hbar^2}{2m} ~(c^\dg \pa_x^2 c ~-~ d^\dg \pa_x^2 d)
\non \\
&& ~~~~~~~~~~- ~\ep \mu ~(c^\dg c ~-~ d^\dg d) \non \\
&& ~~~~~~~~~~+ ~\De ~c^\dg d ~+~ \De ~d^\dg c ], \label{ham6} \eea
where we have chosen the pairing $\De$ to be real. In
Eq.~\eqref{ham6}, $\ep$ is a tuning parameter: for $\ep = 0$, we recover the
Dirac Hamiltonian studied earlier, while for $\ep = 1$, we obtain a
Schr\"odinger Hamiltonian along with a spin-orbit interaction with strength
$\ga$. (In momentum space, the non-superconducting part of the Hamiltonian in
Eq.~\eqref{ham6} is given, in terms of spin-up and spin-down fields $c$ and
$d^\dg$, as $\ep (\hbar^2 k^2 /(2m) - \mu)I + \ga k \si^z$, where $I$ is the
identity matrix).

Given the probability density $\rho = c^\dg c + d^\dg d$, the equations of
motion and continuity imply that the current is
\bea J &=& - ~\frac{i \ep \hbar}{2m} ~(c^\dg \pa_x c ~-~ \pa_x c^\dg c ~-~
d^\dg \pa_x d ~+~ \pa_x d^\dg d) \non \\
&& +~ \ga ~(c^\dg c ~-~ d^\dg d). \label{curr1} \eea

For a semi-infinite system which goes from $x=0$ to $\infty$, we have to impose
the condition $J=0$ at $x=0$ for all the modes. For $\ep = 0$, we saw above
that the general condition which gives zero current at $x=0$ is $c = e^{i \ta}
d$. However, for $\ep = 1$ and $\ga = 0$, we know that the usual condition at
a hard wall is given by $c=0$ and $d=0$. This is not the most general possible
condition which gives zero current for the Schr\"odinger
Hamiltonian~\cite{carreau,harrison}.
However we always require two conditions unlike the case of the Dirac
Hamiltonian where we need only one condition ($c = e^{i \ta} d$).
When both $\ep$ and $\ga$ are non-zero, it is therefore not obvious what
condition should be imposed on $c$, $d$ and their derivatives at $x=0$.

%We now see that even if $\ep \ne 0$, the same condition will give $J=0$.
%Note that this is quite different from the condition $c=d=0$ at $x=0$ that is
%usually chosen at a hard wall for the Schr\"odinger Hamiltonian corresponding
%to $\ep = 1$.

%Krishnendu had suggested starting from a non-relativistic Hamiltonian
%with a spin-orbit term with coupling $\al$,
%\beq H ~=~ \left( \frac{p^2}{2m} ~-~ \mu \right) ~+~ \al p \si^z.
%\label{ham7} \eeq
%This would be a more general starting point, and in an appropriate limit,
%this should give the spin-orbit coupled Dirac Hamiltonian that we have used
%in the earlier sections. We tried to do this but ran into the problem of
%what boundary condition to impose at a junction
%between two systems (which we need to do for the Josephson junction problem).
%The difficulty is that the current has the form
%\beq J ~=~ -~ \frac{i\hbar}{2m} ~(\psi^\dg \pa_x \psi ~-~ \pa_x
%\psi^\dg \psi) ~+~ \al ~\psi^\dg \si^z \psi, \label{curr2} \eeq
%which has terms with and without derivatives. So it was not clear to us what
%matching conditions should be imposed on $\psi$ and $\pa_x \psi$ at
%$x = 0+$ and $0-$ to ensure that the current is continuous across $x=0$.

We therefore turn to a lattice version of this model. The Hamiltonian for
such a model is obtained by adding the following
\bea \de H_l &=& - \frac{g}{2} ~\sum_n ~[ c_{n\ua}^{\dg} c_{n+1\ua} ~+~ 
c_{n\da}^{\dg} c_{n+1 \da} ~+~ {\rm H. c.}] \label{ham7} \eea
to the Hamiltonian given in Eq.~\eqref{ham1}. The eigenvalue equation
therefore changes from Eq.~\eqref{ham3} to
\bea && [(\ga \sin k - g \cos k - \mu) \tau^z + (\De_0 + \De_1
\cos k) \tau^x] \begin{pmatrix} \al \\ \beta \end{pmatrix} \non \\
&=& E\begin{pmatrix} \al \\ \beta \end{pmatrix}, \label{ham8} \eea
which gives 
\beq E = \pm \sqrt{(\ga \sin k - g \cos k - \mu)^{2} + (\De_0 + \De_1 
\cos k )^{2}}. \label{en3} \eeq 
We now consider what
happens if the parameters $\ga, ~\mu, ~\De_0$ and $\De_1$ are held
fixed and $g$ is varied. Eq.~\eqref{en3} implies that the energy gap
will be zero if there is a value of $k$ where $\ga \sin k - g \cos
k - \mu = 0$ (which requires $\sqrt{\ga^2 + g^2} > |\mu|$) and 
$\De_0 + \De_1 \cos k = 0$. The second condition requires 
$|\De_0 /\De_1| \le 1$. Using this in the first condition then implies
that the energy gap will be zero at $g=g_\pm$ where
%\beq g ~=~ \frac{\De_1}{2 \De_0} ~\left[ \mu ~\pm~ \ga ~\sqrt{ 1 ~-~ \left(
%\frac{\De_0}{\De_1} \right)^2} \right]. \label{g} \eeq
\beq g_{\pm} ~=~ \frac{\mu \De_1}{\De_0} ~\pm~ \ga ~\sqrt{ \left( 
\frac{\De_1}{\De_0} \right)^2 ~-~ 1}. \label{g} \eeq

Numerically, we find that if $(\De_0 /\De_1)^2 < 1$ and $g$ lies between 
the values $g_\pm$ given in Eq.~\eqref{g}, the system lies in a topologically 
non-trivial phase and there is a zero energy mode at each end of a
finite-sized system. But if $g$ lies outside this range, there are
no end modes. We find that this also agrees with a winding number
calculation. Defining $a(k) = \ga \sin k - g \cos k - \mu$ and
$b(k) = \De_0 + \De_1 \cos k$, we find that the winding number defined in 
Eq.~\eqref{w} is $\pm 1$ if $g_- < g <g_+$ (consistent with a topologically 
non-trivial phase) and is zero outside this range
(giving a topologically trivial phase). The model therefore hosts two
topological transitions between these phases at $g=g_{\pm}$.

Finally, we note that the equations of motion for the model defined above
have the same two symmetries that we discussed in Sec.~\ref{sec2c}.
This explains why the end modes have zero energy.

\section{Josephson effects for two superconducting systems with different
phases}
\label{sec5}

\subsection{Andreev bound states at a Josephson junction}
\label{sec5a}

\begin{figure}[H]
\centering
\includegraphics[scale=0.52]{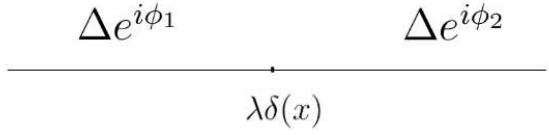}
\caption{Schematic picture of a junction between two $s$-wave superconductors
with pairings $\De e^{i \phi_1}$ ($\De e^{i \phi_2}$) for $x < 0$ ($>0$)
respectively. The junction at $x=0$ has a $\de$-function barrier with
strength $\lm$.} \label{fig06} \end{figure}

In this section, we will study the ABS and the Josephson current between two
superconducting systems in which the $s$-wave pairings have different phases.
We first consider a continuum model. We will take the magnitudes of the two
pairings (and hence the superconducting gaps) to be equal, and their phases to
be $\phi_1$ and $\phi_2$. Further, the two systems will be taken to be
separated by a $\de$-function potential barrier with strength $\lm$ located at
$x=0$. A schematic picture of the system is shown in Fig.~\ref{fig06}. The
continuum Hamiltonians on the two sides of $x=0$ are given by
\bea H_{c1} &=& \int_{-\infty}^0 dx ~[ -~i \ga ~( c^\dg \pa_x c ~-~ d^\dg
\pa_x d) \non \\
&& ~~~~~~~~~~~~~~+~ \De ~e^{i\phi_1} ~c^\dg d ~+~ \De ~e^{-i \phi_1} ~
d^\dg c], \non \\
H_{c2} &=& \int_0^\infty dx ~[ -~i \ga ~( c^\dg \pa_x c ~-~ d^\dg \pa_x
d) \non \\
&& ~~~~~~~~~~~~~~+~ \De ~e^{i\phi_2} ~c^\dg d ~+~ \De ~e^{-i \phi_2} ~
d^\dg c], \label{ham9} \eea
where $H_{c1}$ ($H_{c2}$) is the Hamiltonian on the left (right) of the
$\de$-function barrier respectively.

The equations of motion following from Eqs.~\eqref{ham9}, along with
a time-dependence of $c$ and $d$ of the form $e^{-iEt/\hbar}$, take the form
\bea - i \ga ~\pa_x c ~+~ \De e^{i\phi_i} ~d ~=~ E ~c, \non \\
i \ga ~\pa_x d ~+~ \De e^{-i\phi_i} ~c ~=~ E ~d, \label{eom4} \eea
where $\phi_i = \phi_1 ~(\phi_2)$ for $x < 0 ~(> 0)$ respectively. Complex
conjugating the above equations implies that there is a symmetry under
\beq \phi_i \to \pi - \phi_i ~~~{\rm and}~~~ E \to - E. \label{sym} \eeq

\begin{figure}[H]
\centering
\includegraphics[scale=0.6]{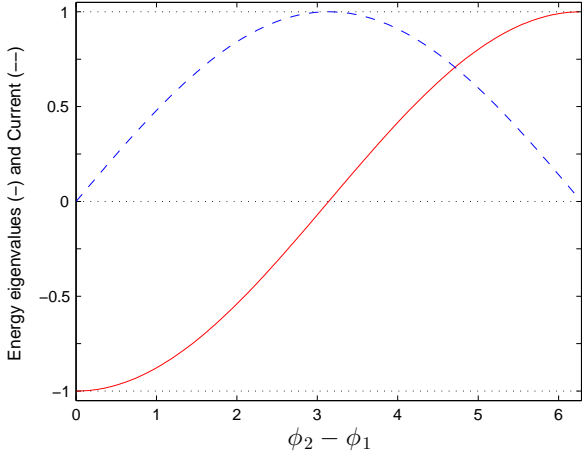}
\caption{Plots of the Andreev bound state energy (red solid curve) in units
of $\De$ and the Josephson current in units of $e \De/\hbar$ (blue dashed 
curve) versus $\phi_2 - \phi_1$, taking $\phi_1 = 0$ and $\ga > 0$.} 
\label{fig07} \end{figure}

Eqs.~\eqref{eom4} imply the energy dispersion $E = \pm \sqrt{\De^2 +
\ga^2 k^2}$, and the second-quantized operators have the form
\bea c_k (x) &=& e^{i(kx - Et)} ~f_k, \non \\
d_k (x) &=& \dfrac{E-\ga k}{\De} ~e^{i(kx - Et) - i \phi} ~f_k. \eea
To find the ABS, the wave number $k$ has to be chosen in such a way that the
wave functions decay away from the $\de$-potential, towards $x \to \pm \infty$ 
on the two sides. From this condition we obtain 
\bea k_1 &=& -~\frac{i}{\ga} ~\sqrt{\De^2 -E^2} ~~~{\rm on ~the ~left}, \non \\
{\rm and} ~~~k_2 &=& \frac{i}{\ga} ~\sqrt{\De^2 - E^2} ~~~{\rm on ~the ~right}.
\label{k12} \eea
The boundary condition at $x=0$ takes the form
\bea c(x=0+) &=& e^{-i \lm /\ga} ~c(x= 0-), \non \\
d(x=0+) &=& e^{-i \lm /\ga} ~d(x=0-). \label{bc} \eea
(We recall that for a Hamiltonian of the Dirac form, a $\de$-function
potential leads to a discontinuity in the wave function. This is unlike
a Hamiltonian of the Schr\"odinger form where a $\de$-function gives
a discontinuity in the first derivative of the wave function).
Since the phase jumps across $x=0$ are equal for $c$ and $d$, we will see that
the $\de$-potential has no effect on expressions for quantities like the
energy spectrum and hence the Josephson current.
% In each region, the wave function has the form
%\begin{center}
%$\begin{pmatrix} c \\ d \end{pmatrix} e^{i(kx - Et)}.$
%\end{center}
Using the boundary condition in Eq.~\eqref{bc}, we can find the
value of the ABS energy. We find that the energy depends only on the
phase difference $\phi_2 - \phi_1$ and has the form 
\beq E ~=~ -~\De ~{\rm sgn} (\ga) ~\cos \left( \dfrac{[\phi_2- \phi_1]}{2} 
\right), \label{abs} \eeq 
where we define the function $[\phi_2 - \phi_1] = \phi_2 - \phi_1$ modulo 
$2 \pi$. Namely, it is a periodic function of $\phi_2 - \phi_1$ with period 
$2 \pi$, and it lies in the range $0 < [\phi_2 - \phi_1] < 2 \pi$. (If 
$\phi_2 - \phi_1$ is exactly equal to a multiple of $2 \pi$, there is, 
strictly speaking, no ABS since such the energy of such a state must satisfy 
$-\De < E < \De$). According to Eq.~\eqref{abs}, when $\phi_2 - \phi_1$ 
approaches a multiple of $2 \pi$, the energy of the ABS approaches $\pm \De$. 
Eqs.~\eqref{k12} then implies that the decay length of the ABS diverges as
$\ga/\sqrt{\De^2 - E^2}$; hence the ABS becomes indistinguishable from the
bulk states. Figure~\ref{fig07} shows the ABS energy $E$ in units of $\De$ 
(red solid curve) as a function of $\phi_2 - \phi_1$, taking $\ga > 0$.
We see that as $\phi_2 - \phi_1$ crosses a 
multiple of $2 \pi$, an ABS disappears after touching the top of the 
superconducting gap and a different ABS appears from the bottom of the gap.

We thus find the peculiar result that there is only one ABS for each
value of $\phi_2 - \phi_1$. One way of understanding why there is
only one ABS instead of two is to note that in our model, there are
only right-moving spin-up and left-moving spin-down electrons. The
ABS is formed by a right-moving spin-up electron which moves from
the left superconductor towards the junction and gets reflected as a
left-moving spin-down hole; alternatively, a left-moving spin-down
electron moves from the right superconductor towards the junction
and gets reflected as a right-moving spin-up hole.

The appearance of a single ABS with the energy given in Eq.~\eqref{abs}
is consistent with a particle-hole transformation in which we transform 
$c \to d$ and $d \to - c$. This transforms our system to a different one 
in which the phases $\phi_1$ and $\phi_2$ have the opposite signs and whose
Hamiltonian also has the opposite sign. Hence all the energy levels (including
the ABS) of the second system should be negative of the energy levels of 
the original system. Indeed we see from Eq.~\eqref{abs} that the sign of
the ABS energy flips when $\phi_i \to - \phi_i$.

We also note that if our sample has a large but finite width, the states 
at the opposite edges would have opposite signs of the velocity $\ga$ in
Eq.~\eqref{ham9}; this is a property of Dirac electrons at the boundaries 
of a topological insulator. If both edges are in proximity to the same 
superconductors so that $\phi_1$ and $\phi_2$ have the same values at the 
two edges, the energies of the ABSs at the two edges will have opposite 
signs. We can see this from Eq.~\eqref{abs} where the expression for the 
energy has a factor of ${\rm sgn} (\ga)$.

%This follows from the fact that $E (-(\phi_2 - \phi_1)) = E (2 \pi - (\phi_2
%- \phi_1))$ which, from Eq.~\eqref{abs}, is equal to $- E (\phi_2 - \phi_1)$.
%The reversal of the energy under $\phi_2 - \phi_1 \to - (\phi_2 - \phi_1)$
%is consistent with the symmetry described in Eq.~\eqref{sym}.

%At zero temperature, the equilibrium Josephson current is given by
%\beq I_{eq} ~=~ \dfrac{2e}{\hbar} \dfrac{\pa E}{\pa (\phi_2 - \phi_1)} \eeq
%if $E < 0$ (the ABS is occupied) and zero if $E > 0$ (the ABS is unoccupied).
%For $\ga > 0$, we therefore obtain
%\bea I_{eq} &=& \dfrac{2e \De}{\hbar} ~\sin \left( \dfrac{\phi_2 - \phi_1}{2}
%\right) ~~~{\rm for}~~~ 0 ~\le~ \phi_2 - \phi_1 ~<~ \pi, \non \\
%&=& 0 ~~~{\rm for}~~~ \pi ~<~ \phi_2 - \phi_1 ~\le~ 2\pi. \eea
%For $2 \pi< \phi_2 < 4\pi$, we have
%\beq I ~=~ -~\dfrac{2e \De}{\hbar} ~\sin \left( \dfrac{\phi_2}{2} \right) \eeq
%which is again positive since $\sin (\phi_2 /2)$ is negative in this interval.
%Thus the Josephson current is always positive. In deriving this, we have
%assumed that the velocity $\ga$ is positive. If we had chosen $\ga < 0$,
%we would have found that the Josephson current is always negative.

Next, we consider the AC Josephson effect. We will consider zero temperature
for simplicity and take $\ga > 0$. If a small constant voltage bias $V_0$ is
applied to the superconductor lying in the region $x > 0$, the pairing phase
there will change slowly in time as
\beq \phi_2 ~=~ \frac{2eV_0 t}{\hbar}. \label{dcjoseph} \eeq
Then the Josephson current will be given by
\bea I_J &=& \dfrac{2e}{\hbar} \dfrac{\pa E}{\pa (\phi_2 - \phi_1)} \non \\
&=& \frac{e \De}{\hbar} ~| \sin \left( \dfrac{\phi_2 - \phi_1}{2} \right) |,
\label{ij1} \eea
where $\phi_2$ changes in time according to Eq.~\eqref{dcjoseph}, and $I_J$ is 
a function of $\phi_2 - \phi_1$ with a periodicity of $2\pi$ as discussed 
after Eq.~\eqref{abs}. Figure~\ref{fig07} shows the Josephson current $I_J$ 
(blue dashed curve) in units of $e \De /\hbar$ as a function of $\phi_2 - 
\phi_1$. Note that $I_J$ has no discontinuity at any value of $\phi_2 - \phi_1$.

Interestingly, we see that $I_J$ is always non-negative, and
therefore its average value (which is also equal to its
time-averaged value since $\phi_2$ varies linearly with time) is
positive. This is unlike the AC Josephson effect found in most
systems where the average value of $I_J$ is zero; hence $I_J$ does
not have a DC part in those systems. Note also that at certain
times, $\phi_2 - \phi_1$ will cross odd-integer multiples of $\pi$;
then the ABS bound state will cross zero energy giving rise to a
fermion-parity switch~\cite{tara}.

We also note as $\phi_2 - \phi_1$ changes in time from zero to $2 \pi$, 
a quasiparticle appears from the bottom of the superconducting gap and moves 
up in energy to reach the top of the gap. Since this quasiparticle
carries spin-up (we recall that both $c^\dg = c_\ua^\dg$ and $d^\dg
= c_\da$ increase the spin component $S^z$ by $\hbar/2$), we have a
process of spin pumping from the left superconductor to the right
superconductor; an amount of $S^z = \hbar/2$ is pumped in a time period 
$2\pi/\om_J$, where $\om_J = 2 e V_0/\hbar$ is the Josephson frequency.

\begin{figure}[H]
\centering
\includegraphics[scale=0.55]{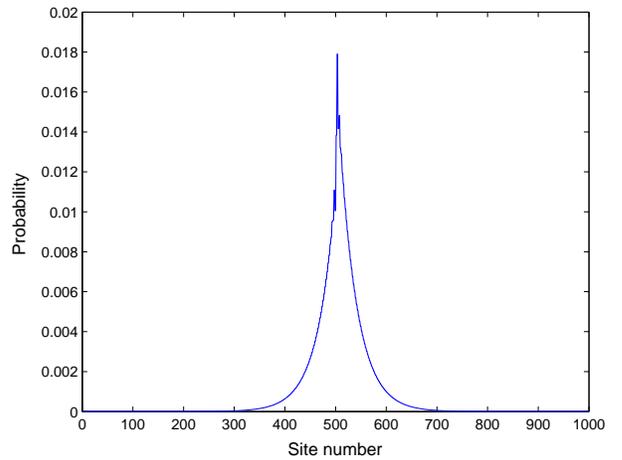}
\caption{Probability $|c_n|^2 + |d_n|^2$ versus $n$ of the Andreev bound
state wave function which appears in the middle of a 500-site system with
pairing $\phi_1 = 0$ in the left half and $\phi_2 = \pi/2$ in the right half
of the system. We have taken $\mu = 0$, $\De_0 = -0.26$, $\De_1 = 0.3$. The 
energy of the Andreev bound state is $-0.028$, and the superconducting gap is 
$0.084$. All energies are in units of $\ga$.} \label{fig08} \end{figure}

We have confirmed the dispersion given in Eq.~\eqref{abs} by doing
numerical calculations for a lattice model. We consider a $500$-site
system with pairing $\phi_1 = 0$ in the left half and $\phi_2 =
\pi/2$ in the right half of the system. We take $\De_0 = -0.26$ and
$\De_1 = 0.3$, so that the pairing of the corresponding continuum
model (given by the modes near $k=0$ of the lattice model) is given
by $\De = \De_0 + \De_1 = 0.04$. We find that there is only one ABS
which lies in the middle of the system; its energy is $-0.028$ which
agrees well with the value of $-\De \cos ((\phi_2 - \phi_1/2)/2)$
given by Eq.~\eqref{abs}. (All energies are in units of $\ga$).
Figure~\ref{fig08} shows the wave function
of this ABS. We have checked numerically that the Fourier transform
of the wave function is sharply peaked around $k=0$ (similar to
Fig.~\ref{fig03} (a)), showing once again that the lattice modes
near $k=\pi$ do not contribute to the ABS. Interestingly, we find
that the expectation value of the charge (Eq.~\eqref{charge}) is
zero in the ABS for any value of its energy.

\subsection{Shapiro plateaus}
\label{sec5b}

In this section, we will study the phenomenon of Shapiro plateaus in a 
resistively and capacitively shunted Josephson junction. In this system, a
resistance $R$ and a capacitance $C$ are placed in parallel with a 
Josephson junction~\cite{ketterson,shukrinov,maiti,erwann,deb}.

Denoting $\phi = \phi_2 - \phi_1$ as the phase difference across the 
Josephson junction, the current across the junction is given by 
Eq.~\eqref{ij1}. The voltage bias across the Josephson junction is given by 
\beq V ~=~ \frac{\hbar}{2e} ~\frac{d \phi}{dt}. \label{vphi} \eeq
The current across the resistance and capacitor are then given by $V/R$
and $C dV/dt$ respectively. The total current is given by the sum of
these three currents. We now impose the condition that the total current
has a constant term $I$ and a term which oscillates with time as $A \sin 
(\om t)$. We thus have the equation 
\beq C \frac{dV}{dt} ~+~ \frac{V}{R} ~+~ \frac{e \De}{\hbar} ~| \sin 
(\frac{\phi}{2})| ~=~ I ~+~ A \sin (\om t). \label{jos1} \eeq
Using Eq.~\eqref{vphi} and introducing the dimensionless time variable
$\tau = \om t$, we can rewrite Eq.~\eqref{jos1} as
\bea \al_1 ~\frac{d^2 \phi}{d \tau^2} ~+~ \al_2 ~\frac{d \phi}{d \tau} ~+~ 
\frac{2 \De}{\hbar \om} ~| \sin (\frac{\phi}{2}) | &=& \frac{2}{e \om} ~[ 
I ~+~ A \sin \tau], \non \\
{\rm where}~~~~ \al_1 ~=~ \frac{\hbar \om C}{e^2} ~~~~{\rm and}~~~~ \al_2 &=&
\frac{\hbar}{e^2 R}. \label{jos2} \eea
We will solve Eq.~\eqref{jos2} numerically from $\tau = 0$ to $\tau = \tau_0$
where $\tau_0$ is a large number (say, $200 \pi$ corresponding to 100 driving
cycles of the current) and find the average value of the voltage bias
\bea \la ~V ~\ra &=& \frac{\hbar \om}{2e} ~\la ~\frac{d \phi}{d \tau} ~\ra 
\non \\
&=& \frac{\hbar \om}{2e}~ \frac{\phi (\tau_0) ~-~ \phi (0)}{\tau_0}. 
\label{avev} \eea
For large values of $\tau_0$, we find numerically that $\la V \ra$ given by 
Eq.~\eqref{avev} does not depend on the initial values of $\phi$ and $d\phi /
d\tau$ at $\tau = 0$.

We will now provide a qualitative
understanding of why Shapiro plateaus should appear in a plot of $\la V \ra$ 
versus $I$. Let us first set $\De = 0$. Eq.~\eqref{jos2} then has the solution 
\bea \phi &=& \chi_1 \tau ~+~ \chi_2 \sin (\tau + \chi_3) ~+~ \phi_0, \non \\
{\rm where} ~~~\chi_1 &=& \frac{2I}{e \om \al_2}, \non \\
\chi_2 &=& - ~\frac{2A}{e \om \sqrt{\al_1^2 ~+~ \al_2^2}}, \non \\
\chi_3 &=& \tan^{-1} \left( \frac{\chi_2}{\chi_1} \right), \label{phi1} \eea
where $\phi_0$ is a constant of integration. The first term in Eq.~\eqref{phi1}
along with Eq.~\eqref{vphi} means that the voltage bias will have an 
average value given by
\beq \la V \ra ~=~ \frac{\hbar \om \chi_1}{2e}. \label{avev2} \eeq
We now substitute Eq.~\eqref{phi1} 
in the third term of the right hand side of Eq.~\eqref{jos2}.
(This procedure can be justified perturbatively if $\De$ is a small parameter).
At this point it is useful to do a Fourier transform of the function 
$| \sin (\pi /2)|$. The Fourier components are given by
\bea F_m &=& \int_0^{2\pi} \frac{d\phi}{2\pi} ~e^{-im\phi} ~|\sin (\phi/2)|
\non \\
&=& \frac{2}{\pi} ~\frac{1}{1 ~-~ 4 m^2}. \label{ijm} \eea
Note that $F_m = F_{-m}$ is real. The third term in Eq.~\eqref{jos2} then
takes the form 
\bea && \frac{2 \De}{\hbar \om} \sum_{m=-\infty}^\infty F_m ~e^{im (\chi_1 
\tau ~+~ \chi_2 \sin (\tau + \chi_3) ~+~ \phi_0)} \non \\
&=& \frac{2 \De}{\hbar \om} \sum_{m=-\infty}^\infty F_m e^{im (\chi_1 
\tau ~+~ \phi_0)} ~\sum_{n=-\infty}^\infty J_n (m \chi_2) ~e^{i n (\tau + 
\chi_3)}, \non \\
&& \label{phi2} \eea
where we have used Eq.~\eqref{phi1} to substitute for $\phi$ in the first 
line, and the Bessel functions in the second line satisfy $J_{-n} (z) = 
J_n (-z) = (-1)^n J_n (z)$.~\cite{abram}. We now see that the expression
in Eq.~\eqref{phi2} will have a DC part which does not vary with $\tau$ 
whenever 
\beq \chi_1 ~=~ - ~\frac{n}{m}, \label{chi1} \eeq
where $m, ~n$ are integers, and we assume that $m \ne 0$.
The corresponding DC part in Eq.~\eqref{phi2} is equivalent to shifting the 
constant $I$ on the right hand side of Eq.~\eqref{jos2} by 
\bea && \frac{e \De}{\hbar} ~[F_m J_n (m \chi_2) e^{i m \phi_0} ~+~ F_{-m}
J_{-n} (-m \chi_2) e^{-i m \phi_0}] \non \\
&=& \frac{2e \De}{\hbar} ~F_m J_n (m \chi_2) \cos (m \phi_0). \label{curr3}
\eea
Since Eq.~\eqref{curr3} can have a range of values depending on $\phi_0$
(since $\cos (m \phi_0)$ can vary from $-1$ to $+1$),
we see that $I$ can have a range of values given by 
\beq \frac{4 e \De}{\hbar} ~F_m J_n (m \chi_2). \label{curr4} \eeq
For all these values of $I$, we see from Eqs.~\eqref{avev2} and \eqref{chi1}
that $\la V \ra$ will have a fixed value given by
\beq \la V \ra ~=~ - ~\frac{\hbar \om}{2 e} ~\frac{n}{m}. \label{avev3} \eeq
This explains why there should be a plateau in $\la V \ra$ for a range of
values of $I$, whenever Eq.~\eqref{avev3} is satisfied. The width of the 
Shapiro plateau will be proportional to $(4 e \De /\hbar) F_m J_n (m \chi_2)$.
Hence the plateau widths go to zero rapidly as either $m$ or $n$ increases 
since $F_m$ goes to zero as $1/m^2$ as $m \to \infty$ and $J_n (z)$ goes to
zero as $(ez/2n)^n$ as $n \to \infty$ keeping $z$ fixed.

We would like to mention here that the series of plateaus corresponding 
to $m=1,2,3,\cdots$ in Eq.~\eqref{avev3} has no analog in standard Josephson 
junctions where $I_J \propto \sin \phi$, and the Fourier transform of $I_J$ 
is non-zero only for $m = \pm 1$. We note that the appearance of such plateaus 
for rational fractional values of $\om_J/\om$ has been noted in a different 
context in Ref.\ \onlinecite{rg1}.

\begin{figure}[H]
\centering
\includegraphics[scale=0.57]{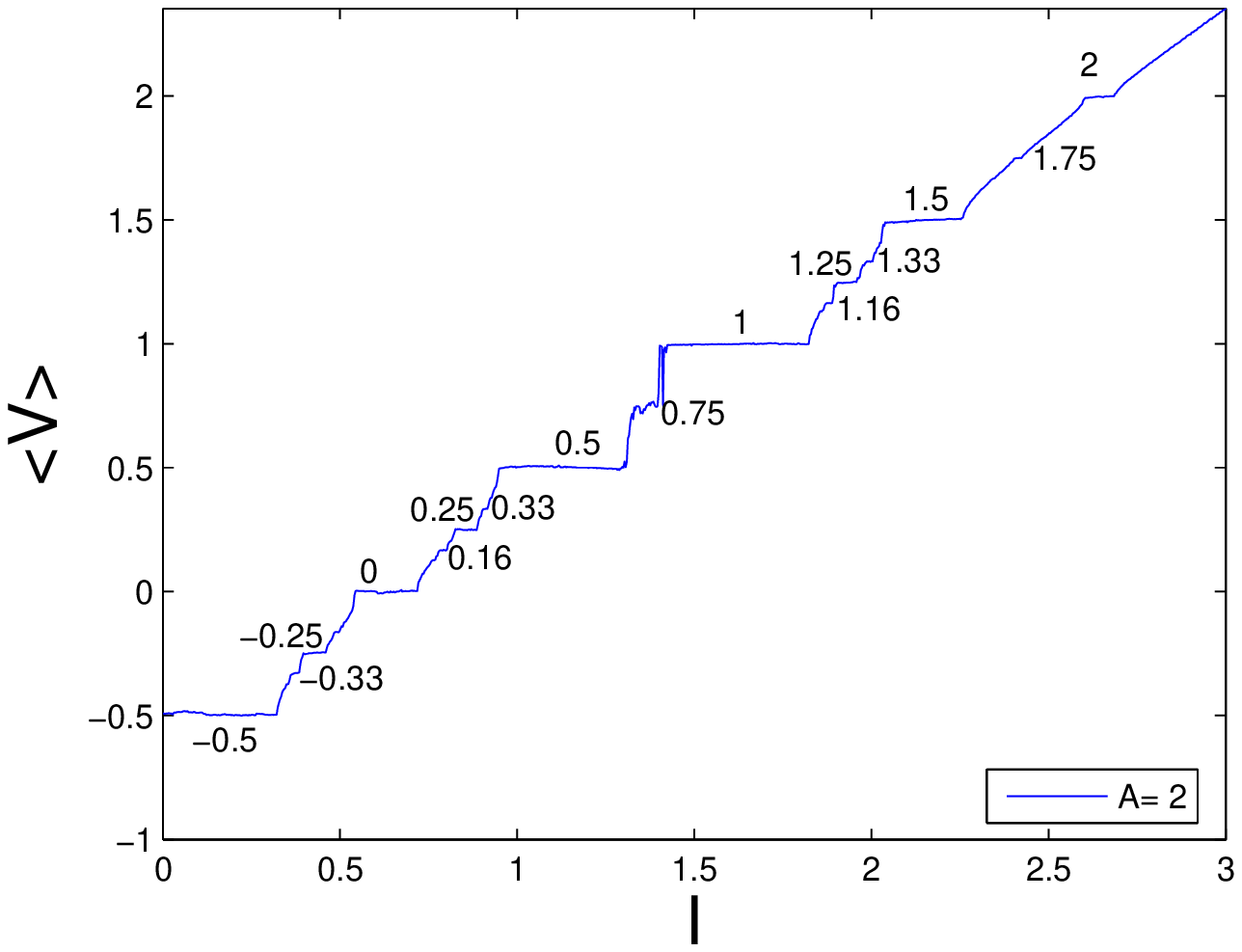}
\caption{Plot of $\la V \ra$ versus $I$. We have chosen $\al_1 = \al_2 = 
1$, $\De = 100 ~\mu$eV, $A=2$, and $\hbar \om = 100 ~\mu$eV. $I$ and $A$ are in
units of $e \om = 24$ nA, and $V$ is in units of $\hbar \om /e = 100 ~\mu$V. 
We see that $\la V \ra$ has several plateaus of different widths at fractional 
multiples of $\hbar \om /(2e)$. The midpoint of the plateau at $\la V \ra = 0$ 
lies at about $I = 2/\pi$. See text for details.} \label{fig09} \end{figure}

We now present our numerical results. For our calculations, we 
choose the parameters in Eq.~\eqref{jos2} as follows:
$\hbar \om = 100 ~\mu$eV which implies $\om \simeq 152$ GHz, $\al_1 = 1$
implying $C \simeq 1.6 \times 10^{-3}$ pF, $\al_2 = 1$ implying $R \simeq 
4.11$ k$\Omega$, and $\De = 100 ~\mu$eV. (This value of the induced 
superconducting gap $\De$ is appropriate for a NbSe$_2$/Bi$_2$Se$_3$ 
superconductor-topological insulator heterostructure~\cite{dai}).
Eq.~\eqref{jos2} then takes the form
\beq \frac{d^2 \phi}{d \tau^2} ~+~ \frac{d \phi}{d \tau} ~+~ 2 ~| \sin 
(\frac{\phi}{2}) | ~=~ \frac{2}{e \om} ~[ I ~+~ A \sin \tau]. \label{jos3} \eeq In Fig.~\ref{fig09}, we show a plot of $\la V \ra$ versus $I$ for $A=2$. 
We note that $I$ and $A$ are given in units of $e \om = 24$ nA, and $\la V
\ra$ is in units of $\hbar \om /e = 100 ~\mu$V. Fig.~\ref{fig09} shows several
plateaus in $\la V \ra$. The most prominent plateaus occur for $\la V \ra$ 
equal to multiples of $\hbar \om /(2e)$ corresponding to the denominator
$m$ in Eq.~\eqref{avev3} being equal to 1. But some small plateaus are also
visible at multiples of $\hbar \om /(4e)$, $\hbar \om /(3e)$, and $\hbar \om 
/(6e)$ corresponding to $m=2$ and $n$ odd, $m=3$ and $n$ even, and $m=3$ and 
$n$ odd in Eq.~\eqref{avev3}.

Fig.~\ref{fig09} shows that the plateau at $\la V \ra = 0$ occurs at 
non-zero values of $I$. This happens because the time-averaged
value of the left hand side of Eq.~\eqref{jos3} is given by the average of 
$2|\sin (\phi/2)|$ over one cycle of $\phi$ from 0 to $2\pi$ which is equal 
to $4/\pi$. This means that the time-averaged value of the right hand 
side of Eq.~\eqref{jos3} is also $4/\pi$. This explains why the midpoint of 
the plateau at $\la V \ra = 0$ lies at about $I/(e \om) = 2/\pi \simeq 0.637$ 
in Fig.~\ref{fig09}. (This is in contrast to other Josephson junctions where 
the current is proportional to $\sin \phi$ or $\sin (\phi /2)$ whose average 
over one cycle is equal to zero). 

Finally, the comment made earlier that a change of $\phi$ by $2\pi$ pumps
an amount of spin angular momentum equal to $\hbar /2$ across the junction
means that the rate of transfer of angular momentum is given by $\hbar /2$
times $d\phi/dt/(2 \pi)$. Eq.~\eqref{vphi} then implies that plateaus in 
$\la V \ra$ are equivalent to plateaus in the average rate of transfer of
angular momentum; the two are related as
\beq \frac{\hbar}{2} ~\frac{\la {d\phi /dt} \ra}{2 \pi} ~=~ \frac{e \la V 
\ra}{2 \pi}. \label{angmomrate} \eeq

\section{Discussion}
\label{sec6}

In this paper we have presented a minimal model of a TRITOPS using the 
chirality of Dirac electrons on a thin, effectively one-dimensional, strip of 
a topological insulator surface. Our model is time-reversal invariant but it
% uses induced extended $s$-wave superconducting pairing and 
possesses half the number of modes of the more well-studied 
time-reversal-invariant topological superconductors due to the chirality of 
the underlying Dirac electrons. This property leads to several unconventional 
features which we have charted out.

We first consider a lattice model of a spin-orbit coupled massless Dirac 
electron in one dimension with $s$-wave superconducting pairing. We 
analytically find the bulk energy spectrum, and use a topological invariant 
called the winding number to identify the regimes of parameter values where 
the system is in topologically trivial and non-trivial phases. In the 
topologically non-trivial phase, a finite-sized system has a single zero 
energy Majorana 
mode at each end; we find that this requires the $s$-wave pairing to have an 
extended form with the magnitude of the nearest-neighbor pairing being larger 
than that of the on-site pairing. For a particular choice of parameters,
we present an analytical expression for the wave function of an end
mode. Although a lattice model of massless Dirac
electrons may suffer from a fermion doubling problem, we find that this
can be avoided in our model if we take the on-site and nearest-neighbor
$s$-wave pairings to have opposite signs and close to each other in magnitude.
Then the wave functions of both the bulk states lying near the gap and
the end modes have momentum components close to $k=0$ rather than
$k=\pi$. The modes near $k=0$ have a smooth continuum limit.

We study the symmetries of the lattice model if both the $s$-wave
pairings are real. These symmetries imply that if there is only one mode at
each end, it must have zero energy and the expectation value of the charge in
such mode will be zero; this is in agreement with our numerical results.
% Hence the end modes have a Majorana character. 
We then consider the effect of making the on-site pairing complex. We find 
that this shifts the energies of the end modes away from zero, but the 
expectation value of the charge remains zero.

We then consider a continuum version of the model with a completely local
$s$-wave pairing. If the pairing is real, this model always turns out to have
zero energy modes at the ends of a long system. The ratio of the phases of the
spin-up electron and spin-down hole wave functions is either $+i$ or $-i$ for
the end modes, and this is found to be in agreement with the lattice results.

Next, we study a lattice system whose Hamiltonian is a combination
of a Schr\"odinger Hamiltonian and a spin-orbit coupled Dirac
Hamiltonian, along with a local $s$-wave pairing. We find that this
system is necessarily topologically trivial if the Dirac part is absent
and can be topologically non-trivial if the Schr\"odinger part is absent. We
analytically find the parameter values at which a topological
transition occurs from one phase to the other. It is worth noting
that an external magnetic field is not required to generate end
modes in any of our models, either on the lattice or in the continuum.

We then study a Josephson junction of two continuum systems which have 
different phases of the $s$-wave pairing, called $\phi_1$ and $\phi_2$. In 
contrast to the earlier models of TRITOPS, we find that there is a single ABS 
which is localized near the junction; its energy depends on the phase 
difference $\De \phi = \phi_2 - \phi_1$ with a period $2\pi$, but it does not 
depend on the strength of a potential barrier which may be present at the
junction (this is related to the Dirac nature of the electrons which
imposes matching conditions on the electron and hole wave functions
but not on their derivatives). As $\De \phi$ varies from 0 to $2 \pi$, the 
ABS energy goes smoothly from the bottom of the superconducting gap to the top.
We then study some Josephson effects at zero temperature. First, we examine 
the AC Josephson effect where a time-independent voltage bias $V_0$ is
applied across the junction. Since this makes $\De \phi$ change
linearly in time, an ABS which initially has negative energy (and is
therefore filled) moves smoothly to positive energy values; this
process repeats periodically in time. We therefore find that the
Josephson current, which is given by the derivative of the ABS
energy with respect to $\De \phi$, varies periodically in time with
a frequency given by $\om_J = 2 e V_0/\hbar$. The Josephson current
turns out be a continuous function of $\De \phi$. However, its sign
does not change with $\De \phi$ which implies that the current has a
non-zero DC component; this is in contrast to the AC Josephson
effect studied earlier in other systems. Second, we consider a 
resistively and capacitively shunted Josephson junction and study what
happens when the voltage bias has both a constant term $V_0$ as a
term $V_1 \cos (\om t)$ which oscillates sinusoidally with an
amplitude $V_1$ and a frequency $\om$. We find that the Josephson
current can then exhibit Shapiro plateaus whenever $\om_J$ is a
rational multiple of $\om$, i.e., $\om_J = (n/m) \om$, where $m, ~n$
are integers. However the plateau widths rapidly go to zero as $m$
or $n$ increases; in particular, if $eV_1/(\hbar \om)$ is small,
only the plateaus with $n=1$ and different values of $m$ would be
observable. The presence of such Shapiro plateaus when $\om_J/\om$
is a rational fraction distinguishes these Josephson junctions from
their standard $s$- or $p$-wave counterparts.

We discuss a few platforms on which our model may be experimentally
realized. A bulk insulating three-dimensional topological insulator where 
one of the surfaces has strong finite-size quantization, allows the
formation of one-dimensional Dirac-like bands that propagate along the surface.
Inducing superconducting by proximity effect on {\it one} such
surface with a conventional $s$-wave superconductor may realize our
model and allow the formation of Majorana bound states at the sample edges 
as we discuss here. One-dimensional Dirac-like states may also be trapped
on one-dimensional crystalline defects that naturally occur on van
der Waals bonded three-dimensional topological insulators
such as Bi$_2$Se$_3$~\cite{alpich,kandala}.
Edges between two facets of a bulk crystal of such a material may also host 
such one-dimensional modes. The proximity of such a state to an $s$-wave
superconductor will realize our model. In the context of two-dimensional 
topological insulators, our model may be realized by inducing
superconductivity using proximity effect on one of the edges of the
sample, leaving the other edge non-proximitized. In a Josephson
junction configuration, the existence of one Andreev bound state,
rather than a pair of Andreev bound states as conventionally
observed, is a striking manifestation of our model. Various
experimental methods including tunneling spectroscopy
~\cite{Lee,Pillet}, Josephson spectroscopy~\cite{Bretheau,geresdi}
and circuit quantum electrodynamics schemes~\cite{hays,tosi} may be
used to detect the presence of a ``single'' Andreev bound state. We
further predict that the Josephson supercurrent in such a geometry
is always positive, which can be detected by DC electrical transport. We 
envisage that such experiments are already possible on various two-dimensional 
and three-dimensional topological insulator materials that are currently
known. Such platforms provide an alternate route towards realization
of Majorana bound states that could potentially display large
topological gaps, and exist at zero magnetic field and at higher
temperatures than currently possible.

\vspace*{.8cm}
\centerline{\bf Acknowledgments}
\vspace{.5cm}

We thank the referees for useful suggestions.
A.B. would like to thank MHRD, India for financial support and P. S.
Anil Kumar for experimental work that inspired some of the ideas.
D.S. thanks DST, India for Project No. SR/S2/JCB-44/2010 for financial 
support. K.S. thanks DST for support through INT/RUS/RFBR/P-314.

\end{document}